\documentstyle[12pt,epsf,rotate]{article}

\def\spose#1{\hbox to 0pt{#1\hss}}
\def\lsim{\mathrel{\spose{\lower 3pt\hbox{$\mathchar"218$}}
 \raise 2.0pt\hbox{$\mathchar"13C$}}}
\def\gsim{\mathrel{\spose{\lower 3pt\hbox{$\mathchar"218$}}
 \raise 2.0pt\hbox{$\mathchar"13E$}}}

\begin{document}

\begin{titlepage}

\begin{flushright}
CERN-TH/97-281\\
Extended version\\
hep-ph/9710331
\end{flushright}

\vspace{1.7cm}
\begin{center}
\boldmath
\large\bf
Towards Extractions of the CKM Angle $\gamma$ from\\ 
\vspace{0.2truecm}
$B_{u,d}\to\pi K$ and Untagged $B_s\to K\overline{K}$ Decays
\unboldmath
\end{center}

\vspace{1.1cm}
\begin{center}
Robert Fleischer\\[0.1cm]
{\sl Theory Division, CERN, CH-1211 Geneva 23, Switzerland}
\end{center}

\vspace{1.2cm}
\begin{abstract}
\vspace{0.2cm}\noindent
The decays $B_d\to\pi^\mp K^\pm$ and $B^\pm\to\pi^\pm K$ provide interesting 
constraints on the angle $\gamma$ of the unitarity triangle of the CKM
matrix. It is shown that bounds on $\gamma$ can also be obtained from the 
time evolution of untagged $B_s\to K^+K^-$ and $B_s\to K^0\overline{K^0}$ 
decays, provided the $B_s$ system exhibits a sizeable width difference 
$\Delta\Gamma_s$. A detailed discussion of rescattering processes and 
electroweak penguin effects, which limit the theoretical accuracy of these
constraints, and of methods to control them through experimental data is
given. Moreover, strategies are pointed out to go beyond these bounds by 
relating the $B_{u,d}\to\pi K$ and untagged $B_s\to K\overline{K}$ decays 
through the $SU(3)$ flavour symmetry of strong interactions. If a tagged, 
time-dependent measurement of the $B_s\to K^+K^-$ and 
$B_s\to K^0\overline{K^0}$ modes should become possible, $\gamma$ could 
be determined from the corresponding observables in a way that makes use 
of only the $SU(2)$ isospin symmetry and takes into account rescattering 
effects ``automatically''. The impact of new-physics contributions to 
$B^0_s$--$\overline{B^0_s}$ mixing is also analysed, and interesting 
features arising in such a scenario of physics beyond the Standard Model 
are pointed out.
\end{abstract}

\vfill
\noindent
CERN-TH/97-281\\
Extended version\\
April 1998

\end{titlepage}

\thispagestyle{empty}
\vbox{}
\newpage
 
\setcounter{page}{1}

\section{Introduction}\label{intro}
The determination of the angle $\gamma$ of the usual non-squashed unitarity
triangle \cite{ut} of the Cabibbo--Kobayashi--Maskawa matrix (CKM matrix) 
\cite{ckm} is considered as being very challenging from an experimental point 
of view \cite{rev}. In order to accomplish this ambitious task, the decays 
$B^0_d\to\pi^-K^+$, $B^+\to\pi^+K^0$ and their charge conjugates are very
promising and have received considerable interest in the recent literature 
\cite{PAPIII}--\cite{defan}. These modes, which have recently been observed 
by the CLEO collaboration \cite{cleo}, should allow us to obtain direct 
information on the CKM angle $\gamma$ at future $B$-factories (BaBar, 
BELLE, CLEO III) (for interesting feasibility studies, see 
\cite{groro,wuegai,babar}). At present, there are only experimental results 
available for the following combined branching ratios \cite{cleo}:
\begin{eqnarray}
\mbox{BR}(B_d\to\pi^\mp K^\pm)&\equiv&\frac{1}{2}
\left[\mbox{BR}(B^0_d\to\pi^-K^+)
+\mbox{BR}(\overline{B^0_d}\to\pi^+K^-)\right]\nonumber\\
&=&\left(1.5^{+0.5}_{-0.4}\pm0.1\pm0.1\right)\times10^{-5}\label{BR-neut}\\
\mbox{BR}(B^\pm\to\pi^\pm K)&\equiv&\frac{1}{2}\left[\mbox{BR}(B^+\to\pi^+K^0)
+\mbox{BR}(B^-\to\pi^-\overline{K^0})\right]\nonumber\\
&=&\left(2.3^{+1.1}_{-1.0}\pm0.3\pm0.2\right)\times 10^{-5}.\label{BR-char}
\end{eqnarray}

In order to determine the CKM angle $\gamma$, the separate branching ratios 
for $B^0_d\to\pi^-K^+$, $B^+\to\pi^+K^0$ and their charge conjugates are 
needed, i.e.\ the combined branching ratios (\ref{BR-neut}) and 
(\ref{BR-char}) are not sufficient, and an additional input is required to 
fix the magnitude of a certain decay amplitude $T$, which is usually 
referred to as a ``tree'' amplitude and will be discussed in more detail 
below \cite{PAPIII,groro,wuegai}. Using arguments based on ``factorization'' 
\cite{facto}, one expects that a future theoretical uncertainty of $|T|$ as 
small as ${\cal O}(10\%)$ may be achievable. Since detailed studies show 
that the properly defined amplitude $T$ is actually not just a 
colour-allowed ``tree'' amplitude, where ``factorization'' may work 
reasonably well \cite{bjorken}, but that it receives also contributions from 
penguin and annihilation topologies due to certain rescattering effects 
\cite{defan,bfm}, these expectations appear too optimistic. In any case, 
some model dependence enters in the extracted value of $\gamma$. 
 
As was pointed out in \cite{fm2}, it is in principle possible to obtain 
{\it constraints} on the CKM angle $\gamma$ that do not suffer from a model 
dependence related to $|T|$. To this end, the combined branching ratios 
(\ref{BR-neut}) and (\ref{BR-char}) are sufficient. In general, such
constraints, which take the form
\begin{equation}\label{gamma-bound1}
0^\circ\leq\gamma\leq\gamma_0\quad\lor\quad180^\circ-\gamma_0\leq\gamma
\leq180^\circ,
\end{equation}
depend also on $|T|$. However, if the ratio 
\begin{equation}\label{Def-R}
R\equiv\frac{\mbox{BR}(B_d\to\pi^\mp K^\pm)}{\mbox{BR}(B^\pm\to\pi^\pm K)}
\end{equation}
of the combined $B_{u,d}\to\pi K$ branching ratios (\ref{BR-neut}) and 
(\ref{BR-char}) is found to be smaller than 1 -- its present experimental 
range is $0.65\pm0.40$, so that this may indeed be the case -- the bound 
$\gamma_0$ takes a maximal value, which is given by
\begin{equation}\label{gam-max}
\gamma_0^{\rm max}=\arccos(\sqrt{1-R})\,,
\end{equation} 
and depends only on $R$. The remarkable feature of these constraints is the
fact that they exclude values of $\gamma$ around $90^\circ$, thereby providing
complementary information to the present ``indirect'' range 
\begin{equation}\label{gamma-range}
41^\circ
\mathrel{\hbox{\rlap{\hbox{\lower4pt\hbox{$\sim$}}}\hbox{$<$}}}\gamma
\mathrel{\hbox{\rlap{\hbox{\lower4pt\hbox{$\sim$}}}\hbox{$<$}}}
134^\circ,
\end{equation}
which arises from the usual fits of the unitarity triangle \cite{burasHF97}. 
A detailed study of the implications of these bounds on $\gamma$ for the
determination of the unitarity triangle can be found in \cite{gnps}. Their 
theoretical accuracy is limited by certain rescattering processes and 
electroweak penguin effects, which led to considerable interest in the recent 
literature \cite{gewe}--\cite{my-FSI} (for earlier references, see 
\cite{FSI}). A comprehensive analysis of these effects and strategies to 
control them through additional experimental data has recently been 
performed in \cite{defan}.
 
In this paper, we focus on the modes $B_s\to K^+K^-$ and 
$B_s\to K^0\overline{K^0}$, which are the $B_s$ counterparts of the 
$B_{u,d}\to\pi K$ decays discussed above, where the up and down ``spectator'' 
quarks are replaced by a strange quark. Because of the expected large 
$B^0_s$--$\overline{B^0_s}$ mixing parameter 
$x_s\equiv\Delta M_s/\Gamma_s={\cal O}(20)$, experimental studies of CP 
violation in $B_s$ decays are regarded as being very difficult. In 
particular, an excellent vertex resolution system is required to keep track 
of the rapid oscillatory $\Delta M_st$-terms arising in tagged $B_s$ 
decays. These terms cancel, however, in {\it untagged} $B_s$ decay rates, 
where one does not distinguish between initially present $B^0_s$ and 
$\overline{B^0_s}$ mesons. In that case, the expected sizeable width 
difference $\Delta\Gamma_s\equiv\Gamma_H^{(s)}-\Gamma_L^{(s)}$ between the 
mass eigenstates $B_s^H$ (``heavy'') and $B_s^L$ (``light'') of the $B_s$ 
system, which may be as large as ${\cal O}(20\%)$ of the average $B_s$ 
decay width $\Gamma_s$ \cite{DGamma}, provides an alternative route to 
explore CP violation \cite{dunietz}. Several strategies have recently been 
proposed to extract CKM phases from such untagged $B_s$ decays 
\cite{dunietz}--\cite{fd2} (for a review, see \cite{hawaii}). 

In Ref.~\cite{fd1}, it was pointed out that untagged data samples of the 
modes $B_s\to K^+K^-$ and $B_s\to K^0\overline{K^0}$ allow a determination of 
the CKM angle $\gamma$. As in the case of the $B_{u,d}\to\pi K$ transitions, 
to this end the magnitude of a certain ``tree'' amplitude $T_s$ has to be 
fixed, leading again to some model dependence of the extracted value of 
$\gamma$. The observables of the untagged $B_s\to K^+K^-$ and $B_s\to 
K^0\overline{K^0}$ modes provide, however, also {\it constraints} on $\gamma$,
which do not depend on such an input. Their theoretical accuracy is also 
limited by certain rescattering and electroweak penguin effects, which will be 
investigated by following closely the formalism developed in \cite{defan}.

The outline of this paper is as follows: in Section~\ref{ampl-obs}, we
introduce a parametrization of the $B_{u,d}\to\pi K$ and 
$B_s\to K\overline{K}$ decay amplitudes in terms of ``physical'' quantities, 
and define the relevant observables. In Section~\ref{gamma-strat}, we discuss 
strategies to constrain and determine the CKM angle $\gamma$ with the help of 
these decays. A detailed discussion of rescattering processes and electroweak 
penguin effects, which limit the theoretical accuracy of these strategies, 
is given in Section~\ref{FSI-EWP-effects}. In Section~\ref{SU3}, we point 
out that the methods to extract $\gamma$ from $B_{u,d}\to\pi K$ and 
$B_s\to K\overline{K}$ decays, which require knowledge of $|T|$ and $|T_s|$, 
can be combined by using the $SU(3)$ flavour symmetry of strong interactions.
Following these lines, such an input can be avoided, and $|T|$ and $|T_s|$
can rather be determined as a ``by-product''. In Section~\ref{tagged}, we 
write a few words on a tagged analysis of the $B_s\to K\overline{K}$ modes, 
which would allow an extraction of $\gamma$ by using only the $SU(2)$ 
isospin symmetry of strong interactions, taking into account rescattering
effects ``automatically''. The impact of CP-violating new-physics 
contributions to $B^0_s$--$\overline{B^0_s}$ mixing for the observables of 
the $B_s\to K\overline{K}$ decays is analysed in Section~\ref{BSM}, and 
the conclusions are given in Section~\ref{concl}.

\boldmath
\section{Decay Amplitudes and Observables}\label{ampl-obs}
\unboldmath
The subject of this section is to introduce a general parametrization of
the $B_{u,d}\to\pi K$ and $B_s\to K\overline{K}$ decay amplitudes in terms
of ``physical'' quantities, and to define the relevant observables to obtain
information on the CKM angle $\gamma$.

\subsection{Decay Amplitudes}
The case of the $B_{u,d}\to\pi K$ decays was discussed in detail in 
\cite{defan}. Using the $SU(2)$ isospin symmetry of strong 
interactions to relate QCD penguin topologies with internal top and charm
quark exchanges (for subtleties related to QCD penguin topologies with 
internal up quarks, see \cite{defan,bfm}), the $B^+\to\pi^+K^0$ and 
$B^0_d\to\pi^-K^+$ decay amplitudes can be expressed as follows:
\begin{eqnarray}
A(B^+\to\pi^+K^0)&=&P\label{ampl-char}\\
A(B^0_d\to\pi^-K^+)&=&-\,[P+T+P_{\rm ew}]\label{ampl-neut}\,,
\end{eqnarray}
where the quantity
\begin{equation}\label{P-def}
P=-\left(1-\frac{\lambda^2}{2}\right)\lambda^2A\left[
1+\rho\,e^{i\theta}e^{i\gamma}\right]{\cal P}_{tc}
\end{equation}
with
\begin{equation}\label{rho-def}
\rho\,e^{i\theta}=\frac{\lambda^2R_b}{1-\lambda^2/2}
\left[1-\left(\frac{{\cal P}_{uc}+{\cal A}}{{\cal P}_{tc}}\right)\right]
\end{equation}
and 
\begin{eqnarray}
{\cal P}_{tc}&\equiv&|{\cal P}_{tc}|e^{i\delta_{tc}}=
\left(P_t-P_c\right)+\left(P_{\rm ew}^t-P_{\rm ew}^c\right)\\
{\cal P}_{uc}&\equiv&|{\cal P}_{uc}|e^{i\delta_{uc}}=
\left(P_u-P_c\right)+\left(P_{\rm ew}^u-P_{\rm ew}^c\right)
\end{eqnarray}
is usually referred to as a $\bar b\to\bar s$ ``penguin'' amplitude. 
In these expressions, $P_q$ and $P_{\rm ew}^q$ denote the contributions of
QCD and electroweak penguin topologies with internal $q$ quarks 
$(q\in\{u,c,t\})$ to $B^+\to\pi^+K^0$, respectively, $\delta_{uc}$, 
$\delta_{tc}$, and $\theta$ are CP-conserving strong phases, the amplitude 
${\cal A}$ is due to annihilation processes, and
\begin{equation}
\lambda\equiv|V_{us}|=0.22\,,\quad
A\equiv\frac{1}{\lambda^2}\left|V_{cb}\right|=0.81\pm0.06\,,\quad
R_b\equiv\frac{1}{\lambda}\left|\frac{V_{ub}}{V_{cb}}\right|=0.36\pm0.08
\end{equation}
are the relevant CKM factors, expressed in terms of the Wolfenstein 
parameters \cite{wolf}. The amplitudes $T$ and $P_{\rm ew}$ -- the latter 
is essentially due to electroweak penguins -- can be parametrized as 
follows \cite{defan}:
\begin{equation}\label{T-def}
T\equiv e^{i\delta_T}e^{i\gamma}|T|=\lambda^4A\,R_b\,e^{i\gamma}\left[
\tilde{\cal T}-{\cal A}+\left(\tilde P_u-P_u\right)+\left(\tilde
P_{\rm ew}^u-\tilde P_{\rm ew}^t\right)-\left(P_{\rm ew}^u-
P_{\rm ew}^t\right)\right]
\end{equation}
\begin{equation}\label{Pew-def}
P_{\rm ew}\equiv-\,|P_{\rm ew}|\,e^{i\delta_{\rm ew}}=-\left(1-
\frac{\lambda^2}{2}\right)\lambda^2A\left[\left(\tilde P_{\rm ew}^t-\tilde
P_{\rm ew}^c\right)-\left(P_{\rm ew}^t-P_{\rm ew}^c\right)\right],
\end{equation}
where the tildes have been introduced to distinguish the $B^0_d\to\pi^-K^+$
amplitudes from those contributing to $B^+\to\pi^+K^0$, and $\delta_T$ and
$\delta_{\rm ew}$ are CP-conserving strong phases. In the literature,
$T$ is usually referred to as a colour-allowed $\bar b\to\bar uu\bar s$ 
``tree'' amplitude. This terminology is, however, misleading in this case, 
since $T$ receives actually not only ``tree'' contributions, which are 
described by $\tilde{\cal T}$, but also contributions from penguin and 
annihilation topologies, as can be seen in (\ref{T-def})~\cite{defan,bfm}. 
The expressions for the charge-conjugate decays $B^-\to\pi^-\overline{K^0}$ 
and $\overline{B^0_d}\to\pi^+K^-$ can be obtained straightforwardly from 
(\ref{ampl-char}) and (\ref{ampl-neut}) by performing the substitution 
$\gamma\to-\,\gamma$ in (\ref{P-def}) and (\ref{T-def}). 

The $B^0_s\to K^0\overline{K^0}$ and $B^0_s\to K^+K^-$ decay amplitudes 
take a completely analogous form to (\ref{ampl-char}) and (\ref{ampl-neut}):
\begin{eqnarray}
A(B^0_s\to K^0\overline{K^0})&=&P_s\label{ampl-neuts}\\
A(B^0_s\to K^+K^-)&=&-\,[P_s+T_s+P_{\rm ew}^s]\label{ampl-chars}\,.
\end{eqnarray}
In contrast to $B^+\to\pi^+ K^0$, its $B_s$ counterpart 
$B^0_s\to K^0\overline{K^0}$ does not receive an annihilation amplitude
corresponding to ${\cal A}$, while an ``exchange'' amplitude 
$\tilde{\cal E}_s$ contributes to $B^0_s\to K^+K^-$, which is not present in 
$B^0_d\to\pi^-K^+$. Moreover, in the case of the $B_s$ modes, we have not
only to deal with ``ordinary'' penguin topologies, but also with 
``penguin annihilation'' processes, which we denote, as the authors of 
\cite{ghlr}, generically by the symbol $(PA)$. Consequently, we obtain the 
following expressions for $\rho_s\,e^{i\theta_s}$, $T_s$ and~$P_{\rm ew}^s$:
\begin{equation}\label{rhos-def}
\rho_s\,e^{i\theta_s}=\frac{\lambda^2R_b}{1-\lambda^2/2}
\left[1-\left\{\frac{{\cal P}_{uc}^s+({\cal PA})_{uc}^s}{{\cal P}_{tc}^s+
({\cal PA})_{tc}^s}\right\}\right]
\end{equation}
\begin{eqnarray}
\lefteqn{T_s\equiv e^{i\delta_T^s}e^{i\gamma}|T_s|=\lambda^4A\,R_b\,
e^{i\gamma}\biggl[\tilde{\cal T}_s+\tilde{\cal E}_s+\left\{\tilde P_u^s+
(\widetilde{PA})_u^s-P_u^s-(PA)_u^s\right\}}\label{Ts-def}\\
&&+\left\{\tilde P_{\rm ew}^{u(s)}+(\widetilde{PA})_{\rm ew}^{u(s)}
-\tilde P_{\rm ew}^{t(s)}-(\widetilde{PA})_{\rm ew}^{t(s)}\right\}\,-\,
\left\{P_{\rm ew}^{u(s)}+(PA)_{\rm ew}^{u(s)}-P_{\rm ew}^{t(s)}-
(PA)_{\rm ew}^{t(s)}\right\}\biggr]\nonumber
\end{eqnarray}
\begin{eqnarray}
\lefteqn{P_{\rm ew}^s\equiv-\,|P_{\rm ew}^s|\,e^{i\delta_{\rm ew}^s}=-\left(1-
\frac{\lambda^2}{2}\right)\lambda^2A}\label{Pews-def}\\
&&\times\left[\left\{\tilde P_{\rm ew}^{t(s)}+(\widetilde{PA})_{\rm ew}^{t(s)}-
\tilde P_{\rm ew}^{c(s)}-(\widetilde{PA})_{\rm ew}^{c(s)}\right\}\,-\,
\left\{P_{\rm ew}^{t(s)}+(PA)_{\rm ew}^{t(s)}-P_{\rm ew}^{c(s)}-
(PA)_{\rm ew}^{c(s)}\right\}\right].\nonumber
\end{eqnarray}
In order to distinguish the $B^0_s\to K^0\overline{K^0}$ decay amplitudes 
from those contributing to $B^0_s\to K^+K^-$, we have introduced, as in the 
$B_{u,d}\to\pi K$ case, tildes to label the latter.

The expected hierarchy of the various contributions to the $B_{u,d}\to\pi K$ 
and $B_s\to K\overline{K}$ modes and the impact of rescattering processes
will be discussed in Section~\ref{FSI-EWP-effects}. For the following 
considerations, the amplitude relations given in (\ref{ampl-char}),
(\ref{ampl-neut}) and (\ref{ampl-neuts}), (\ref{ampl-chars}) are of particular
interest. As was pointed out in \cite{defan}, the amplitudes 
$\rho\,e^{i\theta}$, $T$ and $P_{\rm ew}$ are properly defined ``physical'' 
quantities. A similar comment applies to their $B_s$ counterparts.

\subsection{Observables}
In order to obtain information on the CKM angle $\gamma$ from the decays 
$B^+\to\pi^+K^0$, $B^0_d\to\pi^-K^+$ and their charge conjugates, the following
observables play a key role \cite{defan}:
\begin{eqnarray}
R&\equiv&\frac{\mbox{BR}(B_d\to\pi^\mp K^\pm)}{\mbox{BR}
(B^\pm\to\pi^\pm K)}=1\,-\,2\,\frac{r}{w}\,\left[\,\cos\delta\cos\gamma+\rho\,
\cos(\delta-\theta)\,\right]+r^2\nonumber\\
&&+\,2\,\frac{\epsilon}{w}\,\left[\,\cos\Delta+\rho\,\cos(\Delta-\theta)
\cos\gamma\,\right]\,-\,2\,r\,\epsilon\,\cos(\delta-\Delta)\cos\gamma+
\epsilon^2\label{R-exp}
\end{eqnarray}
\begin{eqnarray}
A_0&\equiv&\frac{\mbox{BR}(B^0_d\to\pi^-K^+)-
\mbox{BR}(\overline{B^0_d}\to\pi^+K^-)}{\mbox{BR}(B^+\to\pi^+K^0)+
\mbox{BR}(B^-\to\pi^-\overline{K^0})}=
A_{\rm CP}(B_d\to\pi^\mp K^\pm)\,R\nonumber\\
&&=A_++2\,\frac{r}{w}\,\sin\delta\sin\gamma+2\,r\,\epsilon\sin(\delta-\Delta)
\sin\gamma+
2\,\epsilon\,\frac{\rho}{w}\,\sin(\Delta-\theta)\sin\gamma\,,\label{A0-exp}
\end{eqnarray}
where
\begin{equation}\label{Ap-def}
A_+\equiv\frac{\mbox{BR}(B^+\to\pi^+K^0)-\mbox{BR}(B^-\to\pi^-
\overline{K^0})}{\mbox{BR}(B^+\to\pi^+K^0)+\mbox{BR}(B^-\to\pi^-
\overline{K^0})}=-\,2\,\frac{\rho}{w^2}\,\sin\theta\sin\gamma
\end{equation}
measures direct CP violation in the decay $B^+\to\pi^+K^0$, and
\begin{equation}\label{w-def}
w\equiv\sqrt{1+2\,\rho\,\cos\theta\cos\gamma+\rho^2}\,.
\end{equation}
In (\ref{R-exp}) and (\ref{A0-exp}), we have introduced the CP-conserving 
strong phase differences
\begin{equation}\label{phases-def}
\delta\equiv\delta_T-\delta_{tc}\,,
\quad\Delta\equiv\delta_{\rm ew}-\delta_{tc}\,,
\end{equation}
and the quantities
\begin{equation}\label{r-eps-def}
r\equiv\frac{|T|}{\sqrt{\left\langle|P|^2\right\rangle}}\,,\quad
\epsilon\equiv\frac{|P_{\rm ew}|}{\sqrt{\left\langle|P|^2\right\rangle}}\,,
\end{equation}
where
\begin{equation}\label{Paver}
\left\langle|P|^2\right\rangle\equiv\frac{1}{2}\left(|P|^2+|\overline{P}|^2
\right).
\end{equation}
Note that tiny phase-space effects have been neglected in (\ref{R-exp}) and 
(\ref{A0-exp}) (for a detailed discussion, see \cite{fm2}).

In the case of the modes $B_s\to K^0\overline{K^0}$ and $B_s\to K^+K^-$,
we have to deal with $B_s$ decays into final states, which are eigenstates
of the CP operator. Taking into account the interference effects arising 
from $B^0_s$--$\overline{B^0_s}$ mixing, the time evolution of the 
corresponding untagged decay rates, which are defined by
\begin{equation}\label{untagged}
\Gamma[f(t)]\equiv\Gamma(B_s^0(t)\to f)\,+\,\Gamma(\overline{B^0_s}(t)\to f)\,,
\end{equation}
where $\Gamma(B_s^0(t)\to f)$ and $\Gamma(\overline{B^0_s}(t)\to f)$ denote
the transition rates corresponding to initially, i.e.\ at time $t=0$, present
$B_s^0$ and $\overline{B^0_s}$ mesons, takes the following form \cite{rev}:
\begin{equation}\label{Bsrate}
\Gamma[f(t)]\propto\left[\left(\frac{1+|\xi_f|^2}{2}\right)-\,\mbox{Re}\,
(\xi_f)\right]e^{-\Gamma_L^{(s)} t}+\left[\left(\frac{1+|\xi_f|^2}{2}\right)+
\mbox{Re}\,(\xi_f)\right]e^{-\Gamma_H^{(s)} t}.
\end{equation}
The observable
\begin{equation}\label{xi-def}
\xi_f=-\,\eta_{\rm CP}^f\,e^{-i\phi_{\rm M}^{(s)}}\,\frac{A(\overline{B^0_s}
\to f)}{A(B^0_s\to f)}
\end{equation}
is proportional to the ratio of the unmixed decay amplitudes 
$A(\overline{B^0_s}\to f)$ and $A(B^0_s\to f)$, and $\eta_{\rm CP}^f$ denotes
the CP eigenvalue of the final state $f$. The weak $B^0_s$--$\overline{B^0_s}$
mixing phase $\phi_{\rm M}^{(s)}=2\,\mbox{arg}(V_{ts}^\ast V_{tb})$ is 
negligibly small in the Standard Model. 

If we use the parametrization of the $B_s\to K\overline{K}$ decay amplitudes 
discussed in the previous subsection, we obtain
\begin{eqnarray}
\Gamma[K^0\overline{K^0}(t)]&=&R_L\,e^{-\Gamma_L^{(s)} t}
\,+\,R_H\,e^{-\Gamma_H^{(s)} t}\label{Bk0k0bar}\\
\Gamma[K^+K^-(t)]&=&\Gamma[K^0\overline{K^0}(0)]
\left[\,a\,e^{-\Gamma_L^{(s)} t}
\,+\,b\,e^{-\Gamma_H^{(s)} t}\right].\label{Bkpkm}
\end{eqnarray}
Introducing the phase-space factor
\begin{equation}
{\cal C}=\frac{1}{8\,\pi\,M_{B_s}}\sqrt{1-4
\left(\frac{M_K}{M_{B_s}}\right)^2}\,,
\end{equation}
the $B_s\to K^0\overline{K^0}$ observables are given by
\begin{eqnarray}
R_L&=&\left(1+2\,\rho_s\,\cos\theta_s\cos\gamma+\rho_s^2\cos^2\gamma\right)
\Gamma_0\label{RL}\\
R_H&=&\left(\rho_s^2\sin^2\gamma\right)\Gamma_0\,,\label{RH}
\end{eqnarray}
where 
\begin{equation}
\Gamma_0={\cal C}\left[\left(1-
\frac{\lambda^2}{2}\right)\lambda^2A\left|{\cal P}_{tc}^s+({\cal PA})_{tc}^s
\right|\right]^2.
\end{equation}
The untagged $B_s\to K^0\overline{K^0}$ decay rate allows us to fix 
$\left\langle|P_s|^2\right\rangle$, which is defined in analogy to 
(\ref{Paver}), through
\begin{equation}\label{Psaver}
{\cal C}\left\langle|P_s|^2\right\rangle=\Gamma[K^0\overline{K^0}(0)]=
R_L+R_H\,.
\end{equation}
Consequently, in order to parametrize the untagged $B_s\to K^+K^-$ rate, we 
may introduce quantities $r_s$ and $\epsilon_s$ as in (\ref{r-eps-def}), 
i.e.\
\begin{equation}\label{rs-epss-def}
r_s\equiv\frac{|T_s|}{\sqrt{\left\langle|P_s|^2\right\rangle}}\,,\quad
\epsilon_s\equiv\frac{|P_{\rm ew}^s|}{\sqrt{\left\langle|P_s|^2\right
\rangle}}\,,
\end{equation}
and obtain 
\begin{eqnarray}
\lefteqn{a=1-2\,\frac{r_s}{w_s}\,\left[\,\cos\delta_s\cos\gamma+\rho_s\,
\cos(\delta_s-\theta_s)\cos^2\gamma\,\right]+r_s^2\,\cos^2\gamma}
\label{a-def}\\
&&+\,2\,\frac{\epsilon_s}{w_s}\,\left[\,\cos\Delta_s+\rho_s\,
\cos(\Delta_s-\theta_s)\cos\gamma\,\right]-2\,r_s\,\epsilon_s\,
\cos(\delta_s-\Delta_s)\cos\gamma+\epsilon_s^2-
\frac{\rho_s^2}{w_s^2}\sin^2\gamma\nonumber
\end{eqnarray}
\begin{equation}\label{b-def}
b=\left[r_s^2-2\,\frac{r_s}{w_s}\,\rho_s\,\cos(\delta_s-\theta_s)+
\frac{\rho_s^2}{w_s^2}\right]\sin^2\gamma\,,
\end{equation}
where 
\begin{equation}\label{ws-def}
w_s\equiv\sqrt{1+2\,\rho_s\,\cos\theta_s\cos\gamma+\rho_s^2}
\end{equation}
corresponds to (\ref{w-def}), and the strong phase differences $\delta_s$ 
and $\Delta_s$ are defined in analogy to (\ref{phases-def}). It is 
interesting to note that we have
\begin{equation}\label{Rs-def}
\frac{\Gamma[K^+K^-(0)]}{\Gamma[K^0\overline{K^0}(0)]}\equiv R_s=a+b\,,
\end{equation}
where $R_s$ takes the same form as the ratio $R$ of the combined 
$B_{u,d}\to\pi K$ branching ratios (see (\ref{R-exp})). 

The expressions given in (\ref{R-exp}), (\ref{A0-exp}), as well as those
given in (\ref{RL}), (\ref{RH}) and (\ref{a-def}), (\ref{b-def}) take into 
account both rescattering and electroweak penguin effects in a completely 
general way and make use only of the isospin symmetry of strong interactions.
Before we analyse these effects in detail in Section~\ref{FSI-EWP-effects}, 
let us first turn to the strategies to constrain and determine the CKM 
angle $\gamma$ from these observables.

\boldmath
\section{Probing the CKM Angle $\gamma$}\label{gamma-strat}
\unboldmath
Before focusing on the decays $B_s\to K^+K^-$ and $B_s\to K^0\overline{K^0}$,
let us spend a few words on strategies to obtain information on the CKM
angle $\gamma$ from their $B_{u,d}$ counterparts $B^0_d\to\pi^-K^+$ and
$B^+\to\pi^+K^0$. For a detailed discussion, the reader is referred to
\cite{defan}. 

\boldmath
\subsection{A Brief Look at the Decays $B_d\to\pi^\mp K^\pm$ and 
$B^\pm\to\pi^\pm K$}
\unboldmath
As soon as the $B_{u,d}\to\pi K$ observables $R$ and $A_0$ have been 
measured, contours in the $\gamma\,$--$\,r$ plane can be fixed. Provided $r$, 
i.e.\ the magnitude of the ``tree'' amplitude $T$, can be fixed as well,
a determination of $\gamma$ becomes possible \cite{PAPIII,groro}. The 
value of $\gamma$ extracted this way suffers, however, from some model 
dependence, which is mainly due to the need to determine $r$ by using an 
additional input (other important limitations arise from rescattering and 
electroweak penguin effects, which will be discussed in 
Section~\ref{FSI-EWP-effects}). The authors of \cite{groro,wuegai} came 
to the conclusion that the future theoretical uncertainty of $r$ in the 
$B$-factory era may be as small as ${\cal O}(10\%)$. This expectation is, 
however, based on arguments using ``factorization'' \cite{facto,bjorken}
and, therefore, is probably too optimistic. In particular, $T$ is not just 
a ``tree'' amplitude, as we have already noted, but it receives in addition 
contributions from penguin and annihilation topologies, which may shift 
its value significantly from the ``factorized'' result owing to certain 
final-state interaction effects \cite{defan}. Interestingly, the present 
CLEO data, summarized in (\ref{BR-neut}) and (\ref{BR-char}), favour values 
of $r$ that are significantly larger than those obtained by applying 
``factorization'' \cite{fm2,defan}, yielding
\begin{equation}\label{r-fact}
\left.r\right|_{\rm fact}=0.16\times a_1\times
\left[\frac{|V_{ub}|}{3.2\times10^{-3}}
\right]\times\sqrt{\left[\frac{2.3\times10^{-5}}{\mbox{BR}(B^\pm 
\to \pi^\pm K)}\right]\times\left[\frac{\tau_{B_u}}{1.6\,\mbox{ps}}\right]}\,.
\end{equation}
Here the relevant $B\to\pi$ form factor obtained in the BSW model 
\cite{BSW} has been used and $a_1\approx1$ is the usual phenomenological 
colour factor \cite{ns}. This interesting feature may be the first indication 
of sizeable non-factorizable contributions to $r$. 

As was pointed out in \cite{fm2}, it may, however, be possible to 
constrain the CKM angle $\gamma$ in a way that is independent of $r$, and
therefore does not suffer from the uncertainty related to this quantity. 
If we use the ``pseudo-asymmetry'' $A_0$ to eliminate the strong phase 
$\delta$ in $R$, the resulting function of $r$ takes the following
minimal value \cite{defan}:
\begin{equation}\label{Rmin}
R_{\rm min}=\kappa\,\sin^2\gamma\,+\,
\frac{1}{\kappa}\left(\frac{A_0}{2\,\sin\gamma}\right)^2.
\end{equation}
In this transparent expression, rescattering and electroweak penguin effects 
are included  through the parameter $\kappa$, which is given by 
\begin{equation}
\kappa=\frac{1}{w^2}\left[\,1+2\,(\epsilon\,w)\cos\Delta+
(\epsilon\,w)^2\,\right].
\end{equation}
The constraints on the CKM angle $\gamma$ are related to the fact that
the values of $\gamma$ implying $R_{\rm min}>R_{\rm exp}$, where $R_{\rm exp}$
denotes the experimentally determined value of $R$, are excluded. If we keep 
both $r$ and  $\delta$ as free, ``unknown'' parameters, we obtain 
$R_{\rm min}=\kappa\,\sin^2\gamma$, leading to the ``original''
bound (\ref{gam-max}), which has been derived in \cite{fm2} for the 
special case $\kappa=1$. For values of $R$ as small as 0.65, which is the 
central value of present CLEO data, a large region around $\gamma=90^\circ$ 
is excluded. As soon as a non-vanishing experimental result for $A_0$ has
been established, also an interval around $\gamma=0^\circ$ and $180^\circ$ 
can be ruled out, while the impact on the excluded region around $90^\circ$ 
is rather small \cite{defan}. 

\boldmath
\subsection{A Closer Look at the Decays $B_s\to K^+K^-$ and 
$B_s\to K^0\overline{K^0}$}\label{Bskk-constr}
\unboldmath
The time evolution of the untagged $B_s\to K^0\overline{K^0}$ decay rate
(\ref{Bk0k0bar}) provides -- in addition to the observables $R_L$ and 
$R_H$ -- the overall normalization of the untagged $B_s\to K^+K^-$  
rate (\ref{Bkpkm}), so that the observables $a$ and $b$ given in 
(\ref{a-def}) and (\ref{b-def}) can be determined. If we neglect for
simplicity rescattering and electroweak penguin effects, i.e.\ $\rho_s=
\epsilon_s=0$, we observe that $a$ and $b$ depend on the three ``unknowns''
$r_s$, $\delta_s$ and $\gamma$. Consequently, an additional input is
required to determine these quantities. Since the observable $b$ fixes a
contour in the $\gamma\,$--$\,r_s$ plane through $r_s=\sqrt{b}/|\sin\gamma|$,
the CKM angle $\gamma$ can be extracted by using additional information on 
$r_s$, for instance the ``factorized'' result corresponding to (\ref{r-fact}) 
\cite{fd1}. The observable $a$ then allows the determination of 
$\cos\delta_s$. Needless to note, this approach suffers from a similar model 
dependence as the $B_{u,d}\to\pi K$ strategy sketched in the previous 
subsection. Since the sum of $a$ and $b$ corresponds exactly to the 
ratio $R$ of the combined $B_{u,d}\to\pi K$ branching ratios, similar bounds 
on $\gamma$, which do not depend on $r_s$, can also be obtained from the 
$B_s\to K\overline{K}$ decays. Moreover, a comparison of $R$ and $R_s$ 
provides interesting insights into $SU(3)$ breaking. 

\begin{figure}
\centerline{
\rotate[r]{
\epsfxsize=9.2truecm
\epsffile{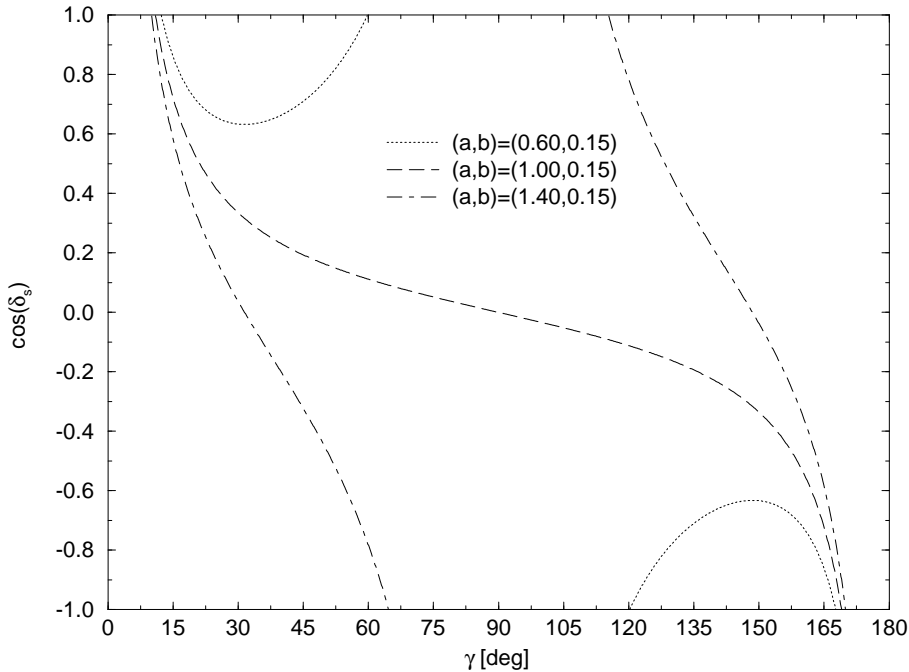}}}
\caption{The contours in the $\gamma\,$--$\,\cos\delta_s$ plane for various 
values of the $B_s\to K^+K^-$ observables $a$ and $b$ in the case of neglected
rescattering and electroweak penguin effects.}\label{fig:CD0}
\end{figure}

A closer look shows, however, that it is possible to derive more elaborate 
bounds from the untagged $B_s\to K\overline{K}$ rates. To this end, 
we use the observable $b$, yielding 
\begin{equation}\label{rs-det}
r_s=\frac{\rho_s}{w_s}\cos(\delta_s-\theta_s)\,\pm\,
\sqrt{\frac{b}{\sin^2\gamma}\,-\,\frac{\rho_s^2}{w_s^2}\sin^2(\delta_s-
\theta_s)}\,,
\end{equation}
to eliminate $r_s$ in the observable $a$ (see (\ref{a-def}) and 
(\ref{b-def})). The resulting expression allows the determination of 
$\cos\delta_s$ as a function of the CKM angle $\gamma$. Since $\cos\delta_s$ 
has to lie within the range between $-\,1$ and $+1$, an allowed range for 
$\gamma$ is implied. Before turning to an analysis of rescattering and 
electroweak penguin effects in the following section, let us here focus again
on the special case $\rho=\epsilon=0$ to illustrate the basic idea of these 
constraints in a transparent way. In this case, we obtain 
\begin{equation}\label{cd0}
\cos\delta_s=\frac{1}{2\sqrt{b}}\left[\left(1-a\right)\tan\gamma\,+\,
\frac{b}{\tan\gamma}\right]\mbox{sgn}(\sin\gamma)\,.
\end{equation}
Since the observable $\varepsilon_K$ measuring indirect CP violation in the
kaon system implies -- using reasonable assumptions about certain hadronic 
parameters -- that $\gamma$ lies within the range $0^\circ\leq\gamma\leq
180^\circ$, we have $\mbox{sgn}(\sin\gamma)=+1$. Let us also note that
estimates based on quark-level calculations indicate $\cos\delta_{(s)}>0$
\cite{fm2,ag}. In Fig.~\ref{fig:CD0}, we show the dependence of 
$\cos\delta_s$ determined with the help of (\ref{cd0}) on $\gamma$ for 
various values of the $B_s\to K^+K^-$ observables $a$ and $b$. The allowed 
regions for $\gamma$ can be read off easily from this figure. They 
correspond to 
\begin{equation}\label{cot-range}
\frac{\left|\,1-\sqrt{a}\,\right|}{\sqrt{b}}\leq|\cot\gamma\,|\leq\frac{1+
\sqrt{a}}{\sqrt{b}}\,,
\end{equation}
and imply
\begin{equation}\label{Bs-bounds}
\gamma_1\leq\gamma\leq\gamma_2\quad\lor\quad180^\circ-\gamma_2\leq\gamma
\leq180^\circ-\gamma_1
\end{equation}
with
\begin{equation}
\gamma_1\equiv\mbox{arccot}\left(\frac{1+\sqrt{a}}{\sqrt{b}}\right)\,,\quad
\gamma_2\equiv\mbox{arccot}\left(\frac{\left|\,1-\sqrt{a}\,\right|}{\sqrt{b}}
\right)\,.
\end{equation}
As a by-product, we get the following bound on the CP-conserving strong 
phase $\delta_s$: 
\begin{equation}
|\cos\delta_s|\geq\sqrt{1-a}\,,
\end{equation}
which becomes non-trivial once $a$ is found experimentally to be smaller 
than 1. It can also be read off nicely from the contours in the
$\gamma\,$--$\,\cos\delta_s$ plane (see Fig.~\ref{fig:CD0}).

In the case of $\rho_s=\epsilon_s=0$, the bounds arising if $R_s$ is measured 
to be smaller than 1, which correspond to those that can be obtained from 
the combined $B_{u,d}\to\pi K$ branching ratios for $R<1$ and are related 
to $R_{\rm min}=\sin^2\gamma$ (see (\ref{gamma-bound1}) and (\ref{gam-max})), 
can be expressed as 
\begin{equation}\label{Rs-bound}
\sqrt{\frac{1-R_s}{R_s}}=\sqrt{\frac{1-a-b}{a+b}}\leq|\cot\gamma\,|\,.
\end{equation}
There are two differences between this bound and the one given in 
(\ref{cot-range}). First, (\ref{cot-range}) allows us to exclude 
$\gamma=0^\circ$ or $180^\circ$. Second, the bound (\ref{cot-range}) is 
more stringent than the $R_s$-bound (\ref{Rs-bound}), i.e.\ excludes a larger 
region around $\gamma=90^\circ$, if we have
\begin{equation}
\gamma_2<\gamma_{0,s}^{\rm max}\,,
\end{equation}
where $\gamma_{0,s}^{\rm max}\equiv\arccos(\sqrt{1-R_s})$. This inequality 
is, however, equivalent to
\begin{equation}
\left(\,a-\sqrt{a}+b\,\right)^2>0\,,
\end{equation}
which is trivially satisfied, unless
\begin{equation}
R_s\equiv a+b=\sqrt{a}\,.
\end{equation}
In that particular case, we have $\gamma_2=\gamma_{0,s}^{\rm max}$, so that
(\ref{cot-range}) and (\ref{Rs-bound}) exclude the same region around 
$\gamma=90^\circ$. Besides a sizeable width difference $\Delta\Gamma_s$ 
and non-vanishing values of $a$ and $b$, the bound (\ref{cot-range}) does 
not require any constraint on these observables such as $a+b<1$, which is 
needed for (\ref{Rs-bound}) to become effective. Unless future experiments 
encounter the special case $a=1$, always a certain range around 
$\gamma=90^\circ$ can be excluded, which is of particular phenomenological 
interest.

\boldmath
\section{Impact of Rescattering Processes and\\ 
Electroweak Penguins}\label{FSI-EWP-effects}
\unboldmath
The issue of rescattering effects in the decays $B^\pm\to\pi^\pm K$ and 
$B_d\to\pi^\mp K^\pm$, originating from processes of the kind 
$B^+\to\{\pi^0K^+,\,\pi^0K^{\ast +},\,\rho^0K^{\ast +},\,\ldots\,\}
\to\pi^+K^0$, led to considerable interest in the recent literature 
\cite{gewe}--\cite{my-FSI}. A detailed study of the
impact of these final-state interaction effects on the strategies to 
constrain and determine the CKM angle $\gamma$ from $B^\pm\to\pi^\pm K$ 
and $B_d\to\pi^\mp K^\pm$ decays, which we briefly discussed in the 
previous section, was performed in \cite{defan}. Concerning the contours 
in the $\gamma\,$--$\,r$ plane and the related bounds on $\gamma$, these 
effects can be taken into account completely by using additional experimental 
information on the decay $B^+\to K^+\overline{K^0}$ and its charge conjugate 
\cite{defan,my-FSI}. Interestingly, the corresponding combined branching 
ratio may be considerably enhanced through rescattering processes to 
the ${\cal O}(10^{-5})$ level, so that $B^\pm\to K^\pm K$ may be measurable 
at future $B$-factories. A detailed analysis of the impact of electroweak 
penguins on the $B_{u,d}\to\pi K$ decays, which were raised in 
\cite{groro,neubert,fm3}, was also performed in \cite{defan}. In this 
section, we will therefore focus on the $B_s$ modes $B_s\to K^0\overline{K^0}$ 
and $B_s\to K^+K^-$.

\subsection{The Role of Rescattering Processes}\label{res-effects}
The parameters $\rho$ and $\rho_s$ are highly CKM-suppressed by 
$\lambda^2R_b\approx0.02$, as can be seen in (\ref{rho-def}) and 
(\ref{rhos-def}). Model calculations performed at the perturbative quark 
level typically give $\rho,\rho_s={\cal O}(1\%)$ and do not indicate a 
significant compensation of this very large CKM suppression. However, 
in a recent attempt \cite{fknp} to evaluate rescattering processes such as
$B^+\to\{\pi^0K^+\}\to\pi^+K^0$, it is found that 
$|{\cal P}_{uc}|/|{\cal P}_{tc}|={\cal O}(5)$, implying that $\rho$ may 
be as large as ${\cal O}(10\%)$. A similar feature arises also in a simple 
model to describe final-state interactions, which assumes elastic 
rescattering processes and has been proposed in \cite{gewe,neubert}. 
An interesting phenomenological implication of large rescattering effects 
would be sizeable CP violation in $B^\pm\to\pi^\pm K$, allowing a first step 
towards constraining the parameter $\rho$ \cite{defan}.

\begin{figure}
\begin{center}
\leavevmode
\vspace*{1truecm} 
\rotate[r]{
\epsfysize=11.8truecm 
\epsffile{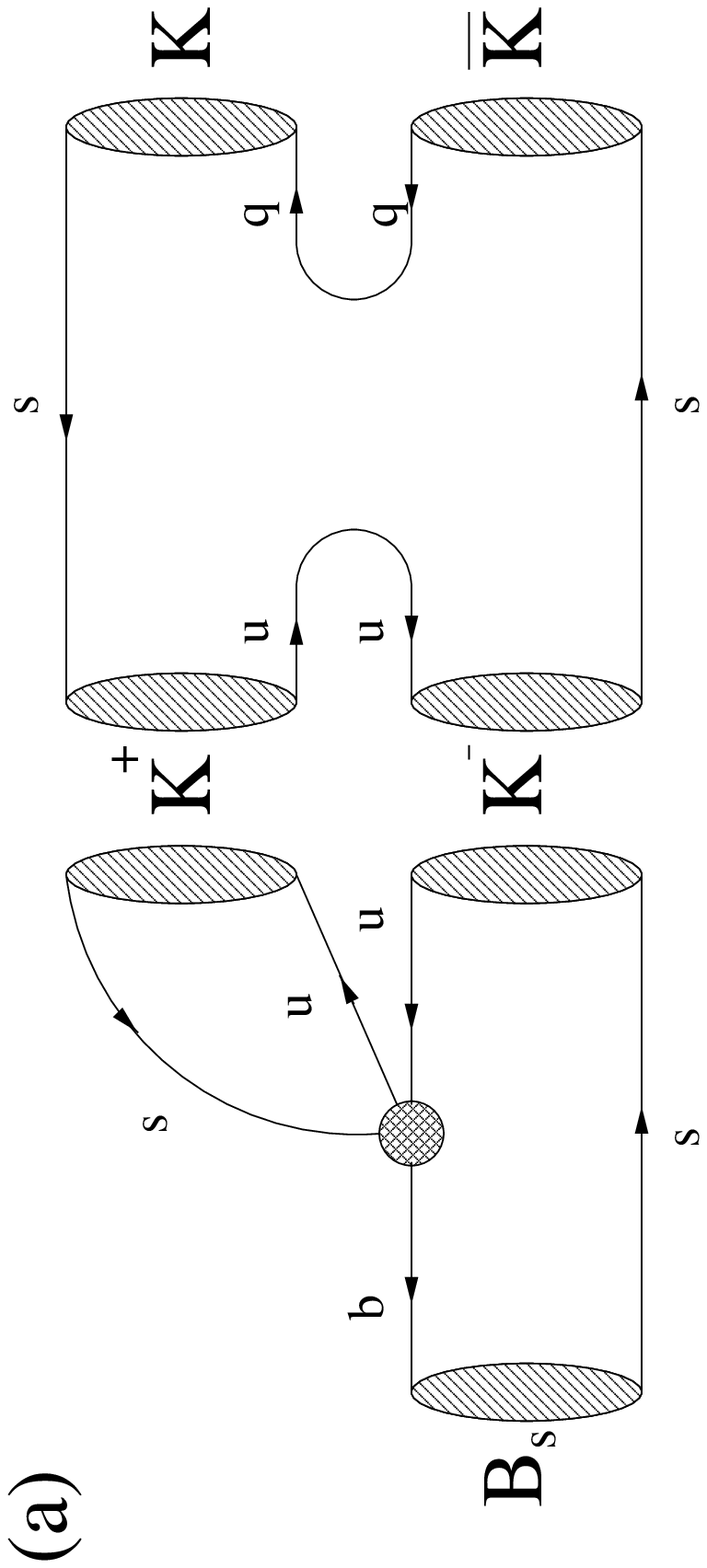}}
\vspace*{1truecm} 
\rotate[r]{
\epsfysize=11.5truecm 
\epsffile{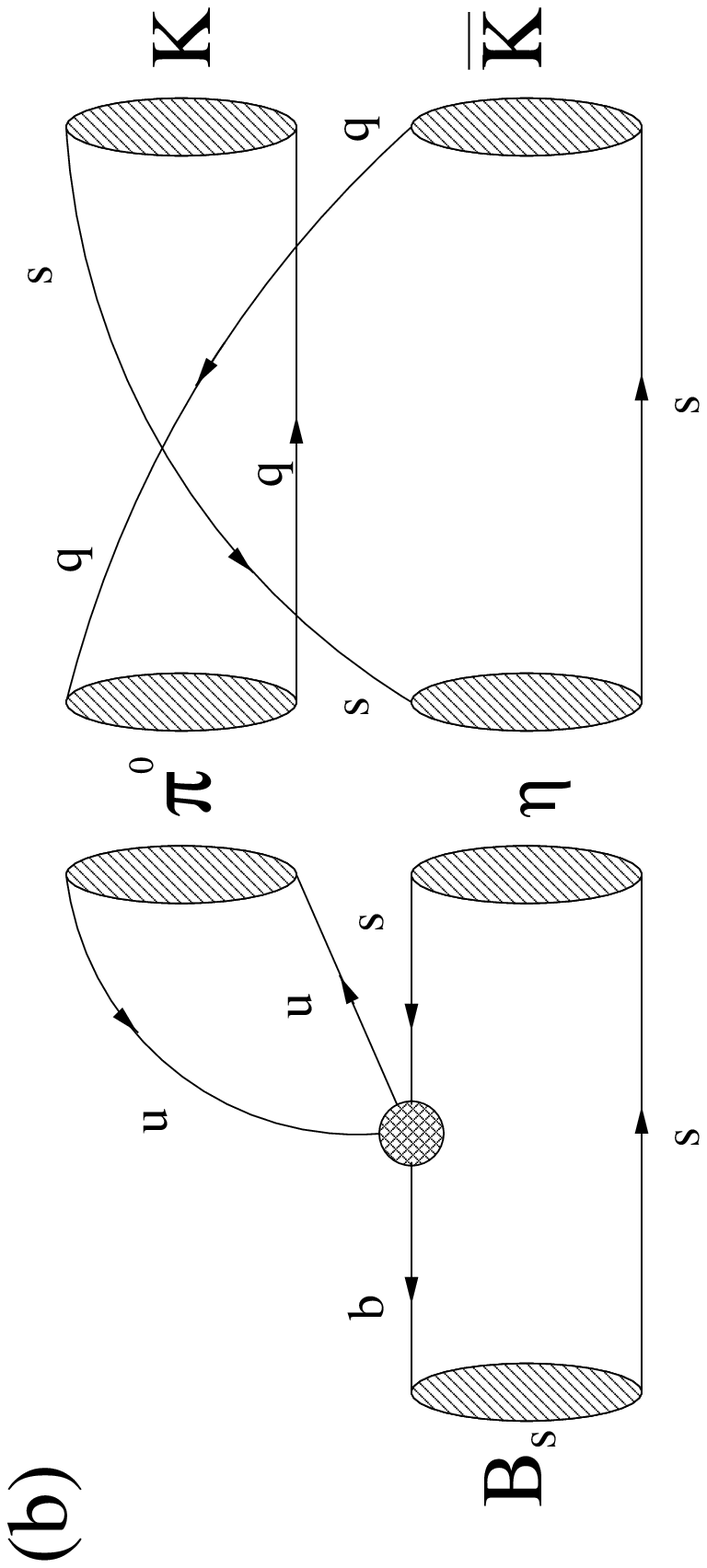}}
\rotate[r]{
\epsfysize=11.3truecm 
\epsffile{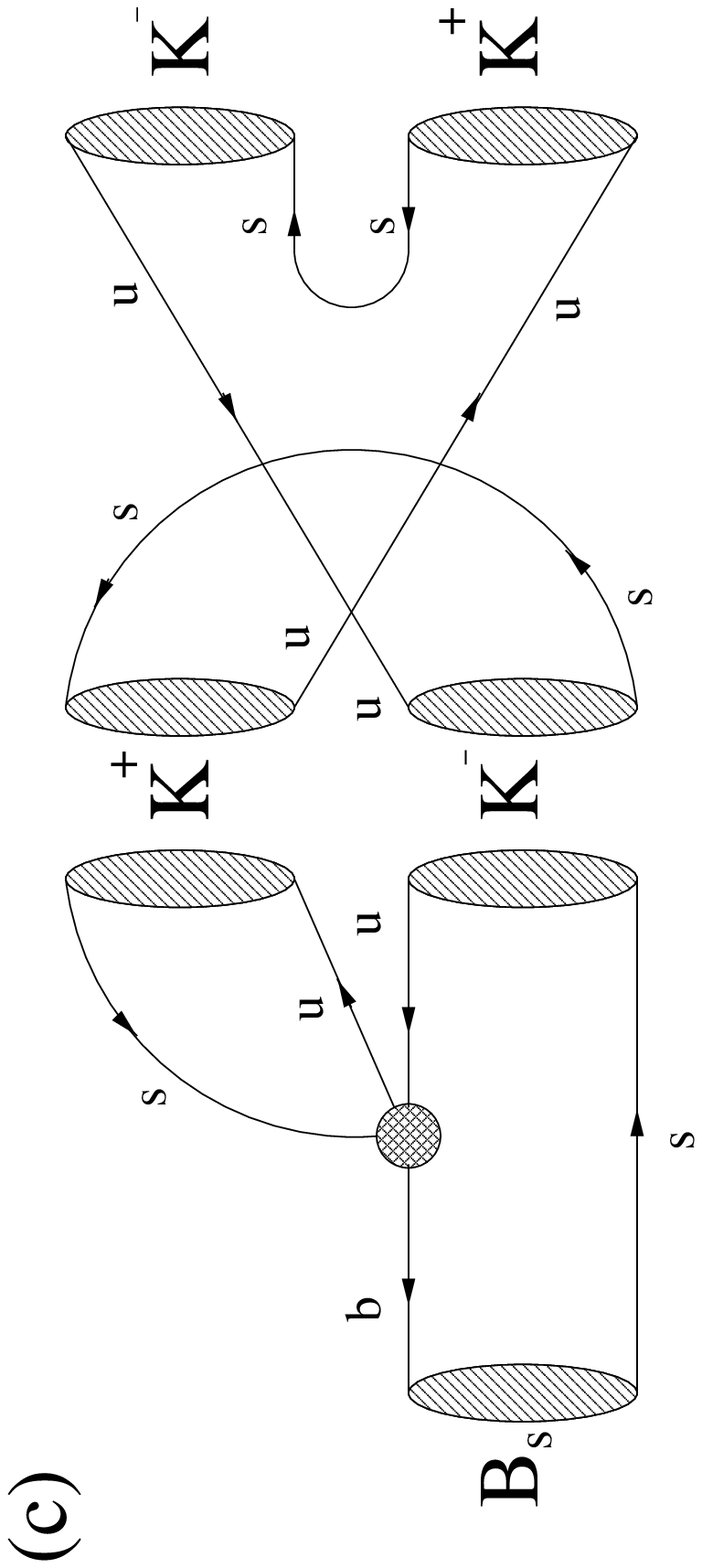}}
\end{center}
\caption{Examples of rescattering processes contributing to $B_s\to K
\overline{K}$. The shaded circles represent insertions of the usual 
current--current operators $Q_{1,2}^u$, and $q\in\{u,d\}$ distinguishes 
between the final states $K^+K^-$ and $K^0\overline{K^0}$. The 
annihilation-like topology (c) arises only in the case of $B^0_s\to K^+K^-$ 
and contributes to the amplitude 
$\tilde{\cal E}_s$.}\label{fig:rescatter-topol}
\end{figure}

As was pointed out in \cite{neubert,Bloketal}, the usual argument for the 
suppression of annihilation processes relative to tree-diagram-like 
topologies by a factor $f_B/m_B$ does not apply to rescattering processes.  
Consequently, these topologies may also play a more important role than 
na\"\i vely expected. Model calculations \cite{Bloketal} based on Regge 
phenomenology typically give an enhancement of the ratio 
$|{\cal A}|/|\tilde{\cal T}|$ from $f_B/m_B\approx0.04$ to ${\cal O}(0.2)$. 
Rescattering processes of this kind can be probed, e.g.\ by the 
$\Delta S$\,=\,0 decay $B^0_d\to K^+K^-$. A future stringent bound on 
BR$(B^0_d\to K^+K^-)$ at the level of $10^{-7}$ or lower may provide a 
useful limit on these rescattering effects~\cite{groro}. The present upper 
bound obtained by the CLEO collaboration is $4.3\times10^{-6}$ \cite{cleo}.
 
In the case of $B_s\to K\overline{K}$ decays, we have to deal with 
final-state interaction effects related to rescattering processes such 
as $B_s\to\{K^+K^-\}\to K^0\overline{K^0}$, which are illustrated in 
Fig.~\ref{fig:rescatter-topol}. Here the shaded circles represent 
insertions of the current--current operators 
\begin{equation}\label{CC-ops}
Q_1^u=(\bar u_{\alpha} s_{\beta})_{{\rm V-A}}\;(\bar b_{\beta} 
u_{\alpha})_{{\rm V-A}}\,,\quad
Q_2^u=(\bar u_{\alpha} s_{\alpha})_{{\rm V-A}}\;
(\bar b_{\beta} u_{\beta})_{{\rm V-A}}\,, 
\end{equation}
where $\alpha$ and $\beta$ are $SU(3)_{\rm C}$ colour indices. The topologies 
of the kind (a) and (b) shown in this figure contribute both to 
$\rho_s\, e^{i\theta_s}$ (see (\ref{rhos-def})), while (c) represents 
a potentially important contribution to the ``exchange'' amplitude 
$\tilde{\cal E}_s$, arising in the case of $B_s\to K^+K^-$. Topologies of 
the latter kind describe also contributions to the ``penguin annihilation'' 
amplitudes $(PA)_s$, if the current--current operators are replaced by
penguin operators.

Although specific model calculations of these $B_s$ rescattering processes 
have not yet been performed, it is plausible to assume that features similar
to those occurring for their $B_{u,d}$ counterparts may show up. Consequently, 
$\rho_s$ may be as large as ${\cal O}(10\%)$, and the role of the exchange 
and penguin annihilation amplitudes may be underestimated by the na\"\i ve 
expectation $|\tilde{\cal E}_s|/|\tilde{\cal T}_s|\approx|(PA)_s|/
|P_s|\approx0.04$. An important experimental tool to investigate the
latter feature is provided by the transition $B_s\to\pi^+\pi^-$, which 
exhibits a decay amplitude proportional to $\tilde{\cal E}_s+(PA)_s$. 
The na\"\i ve expectation for the corresponding branching ratio is 
${\cal O}(10^{-8})$, and a significant enhancement would indicate that 
rescattering contributions, as those shown in Fig.~\ref{fig:rescatter-topol} 
(c), play an important role. 

\begin{figure}
\centerline{
\rotate[r]{
\epsfxsize=9.2truecm
\epsffile{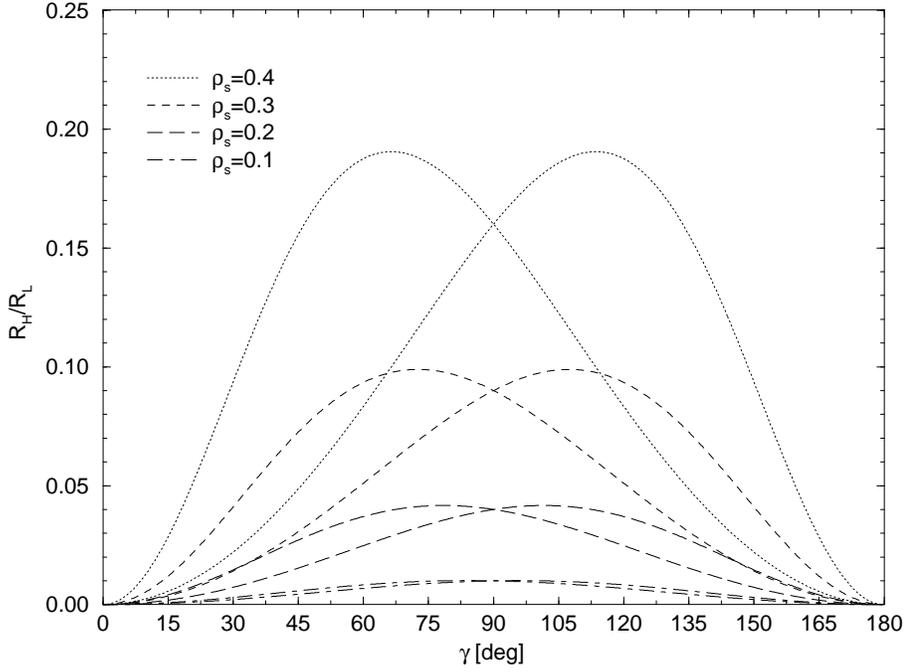}}}
\caption{The dependence of the ratio $R_H/R_L$ of the $B_s\to K^0
\overline{K^0}$ observables on the CKM angle $\gamma$ for various values 
of $\rho_s$ and $\theta_s\in\{0^\circ,180^\circ\}$.}\label{fig:RHRL}
\end{figure}

If we look at the untagged $B_s\to K^0\overline{K^0}$ rate (\ref{Bk0k0bar}) 
and their observables (\ref{RL}) and (\ref{RH}), we notice that the 
term proportional to $R_H$, evolving in time with the decay width 
$\Gamma_H^{(s)}$, is essentially due to rescattering processes. In 
Fig.~\ref{fig:RHRL}, we illustrate this feature, which has some analogy to
the generation of the CP asymmetry $A_+$ arising in $B^\pm\to\pi^\pm K$, 
by showing the dependence of the ratio $R_H/R_L$ on the CKM angle $\gamma$ 
for various values of $\rho_s$ and $\theta_s\in\{0^\circ,180^\circ\}$. 
Since $R_H$ is proportional to $\rho_s^2$, final-state interaction effects 
may lead to values of $R_H/R_L$ of at most ${\cal O}(10\%)$. Consequently, 
a measurement of the part of the untagged $B_s\to K^0\overline{K^0}$ rate 
proportional to $e^{-\Gamma_H^{(s)}t}$ may be very difficult. However, as 
we will see in Section~\ref{BSM}, the ratio $R_H/R_L$ could be dramatically 
enhanced from this Standard Model expectation, if 
$B^0_s$--$\overline{B^0_s}$ mixing receives sizeable CP-violating 
contributions from new physics. 

\begin{figure}
\centerline{
\rotate[r]{
\epsfxsize=9.2truecm
\epsffile{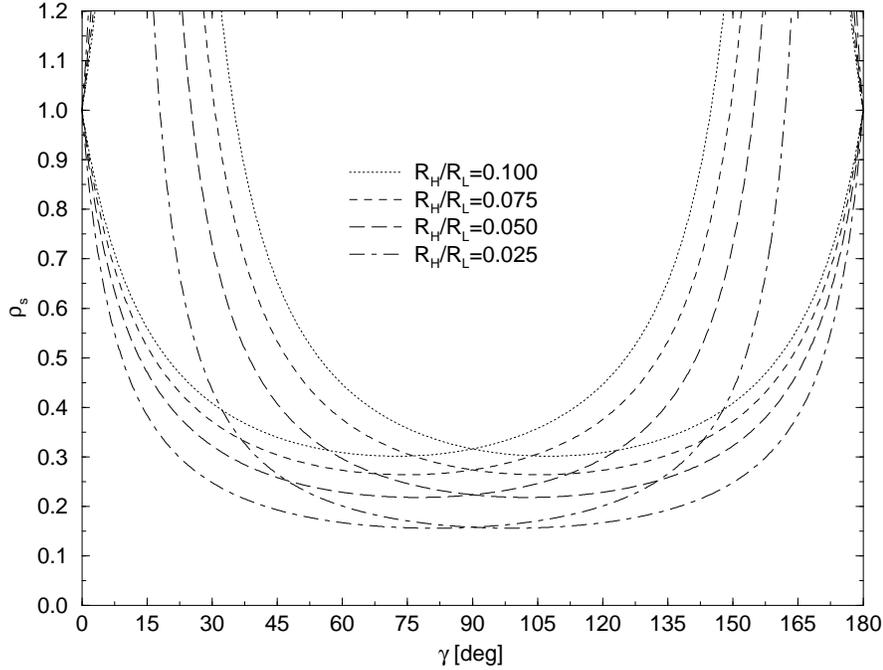}}}
\caption{The constraints on the parameter $\rho_s$ arising from the 
observables $R_H$ and $R_L$ of the untagged $B_s\to K^0\overline{K^0}$ 
rate.}\label{fig:rho-RHRL}
\end{figure}

In the case of the decay $B^+\to\pi^+K^0$, the CP asymmetry $A_+$ implies
upper and lower bounds for $\rho$, which are given by 
\begin{equation}\label{rho-min-max}
\rho_{\rm min}^{\rm max}=\frac{\sqrt{A_+^2\,+\,
\left(1-A_+^2\right)\sin^2\gamma}\,\pm\,\sqrt{\left(1-A_+^2\right)
\sin^2\gamma}}{|A_+|}\,,
\end{equation}
and are discussed in detail in \cite{defan}. Concerning the mode 
$B_s\to K^0\overline{K^0}$, the ratio of the observables $R_H$ and $R_L$ 
implies upper and lower bounds for $\rho_s$, which take the form
\begin{equation}\label{rhos-bounds}
(\rho_s)^{\rm max}_{\rm min}=\frac{\sqrt{R_H}}{\left|\sqrt{R_L}\,
|\sin\gamma|\,\mp\,\sqrt{R_H}\,|\cos\gamma|\right|}\,,
\end{equation}
and are very different from (\ref{rho-min-max}). In Fig.~\ref{fig:rho-RHRL},
we show the dependence of (\ref{rhos-bounds}) on the CKM angle $\gamma$ for
various values of $R_H/R_L$. Interestingly, we have $(\rho_s)_{\rm min}=
(\rho_s)^{\rm max}=\sqrt{R_H/R_L}$ for $\gamma=90^\circ$, so that $\rho_s$ 
is fixed completely in this case. The corresponding CP-conserving strong 
phases $(\theta_s)^{\rm max}_{\rm min}$ are given by 
\begin{equation}\label{thetas-bounds}
(\theta_s)_{\rm min}=\left\{\begin{array}{lcl}180^\circ&\,\mbox{for}\,&
0^\circ<\gamma<90^\circ\\
0^\circ&\,\mbox{for}\,&90^\circ<\gamma<180^\circ
\end{array}\right\},
\,\,\,
(\theta_s)^{\rm max}=\left\{\begin{array}{lcl}0^\circ&\,\mbox{for}\,&
0^\circ<\gamma<90^\circ\\
180^\circ&\,\mbox{for}\,&90^\circ<\gamma<180^\circ
\end{array}\right\}
\end{equation}
for $|\tan\gamma|>\sqrt{R_H/R_L}$, and by $(\theta_s)^{\rm max}=
(\theta_s)_{\rm min}$ with $(\theta_s)_{\rm min}$ given in 
(\ref{thetas-bounds}) otherwise.

\begin{figure}
\centerline{
\rotate[r]{
\epsfxsize=9.2truecm
\epsffile{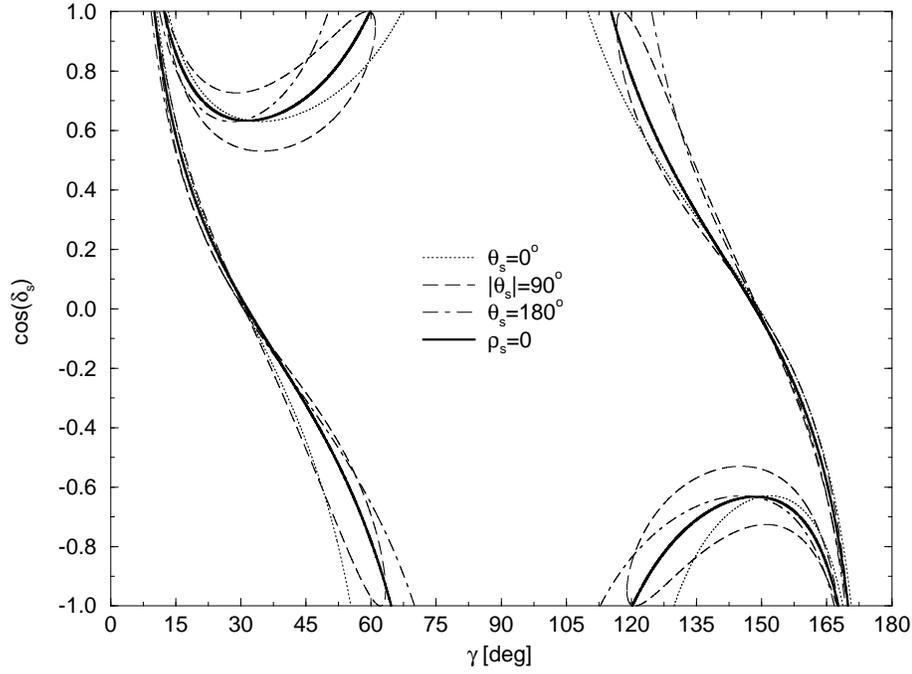}}}
\caption{The contours in the $\gamma\,$--$\,\cos\delta_s$ plane corresponding
to $(a,b)=(0.60,0.15)$ and $(1.40,0.15)$ in the presence of rescattering 
effects described by $\rho_s=0.15$.}
\label{fig:FSI-rho015}
\end{figure}

In order to constrain the CKM angle $\gamma$ in the presence of rescattering
effects, i.e.\ $\rho_s\not=0$, we follow the strategy discussed in 
Subsection~\ref{Bskk-constr} and eliminate $r_s$ in (\ref{a-def}) through
(\ref{rs-det}), which yields the equation
\begin{equation}\label{cent-eq}
A_\rho\,\cos^2\delta_s\,+\,2\,B_\rho\,
\cos\delta_s\,\sin\delta_s\,-\,C^2_\rho\,=\,0\,,
\end{equation}
having the solution
\begin{equation}\label{eq-sol}
\cos^2\delta_s=\frac{A_\rho\,C^2_\rho\,+
\,2\,B^2_\rho\,\pm\,2\,B_\rho\,
\sqrt{A_\rho\,C^2_\rho\,+\,B^2_\rho\,-
\,C^4_\rho}}{A^2_\rho\,+\,4\,B_\rho^2}
\end{equation}
with
\begin{eqnarray}
A_\rho&=&\frac{4}{w_s^2}\left[\left(\frac{b}{\sin^2\gamma}
\right)\,-\,\frac{\rho_s^2}{w_s^2}\right]\cos^2\gamma\,+
\,4\,C_\rho\frac{\rho_s}{w_s^2}
\cos\theta_s\,\cos\gamma\label{A-Def}\\
B_\rho&=&2\,C_\rho\,\frac{\rho_s}{w_s^2}\,
\sin\theta_s\,\cos\gamma\\
C_\rho&=&1\,-\,a\,+\,b\left(\frac{\cos\gamma}{\sin\gamma}
\right)^2-\,\frac{\rho_s^2}{w_s^2}\,.\label{C-Def}
\end{eqnarray}
In these expressions, which represent the generalization of (\ref{cd0})
in the presence of rescattering processes, electroweak penguin effects 
have been neglected, i.e.\ $\epsilon_s=0$. The corresponding contours in 
the $\gamma\,$--$\,\cos\delta_s$ plane are shown in Fig.~\ref{fig:FSI-rho015}
for $\rho_s=0.15$, various values of the strong phase $\theta_s$, and 
the $B_s\to K^+K^-$ observables $(a,b)=(0.60,0.15)$ and $(1.40,0.15)$. 
The thick solid lines represent the curves shown in Fig.~\ref{fig:CD0}. 
We observe that the final-state interaction effects are negligible in
the case of $\theta_s=\pm\,90^\circ$, and are maximal for $\theta_s\in
\{0^\circ,180^\circ\}$, leading to an uncertainty of $\Delta\gamma=\pm\,
8^\circ$ in this specific example. 

\begin{figure}
\centerline{
\rotate[r]{
\epsfxsize=9.2truecm
\epsffile{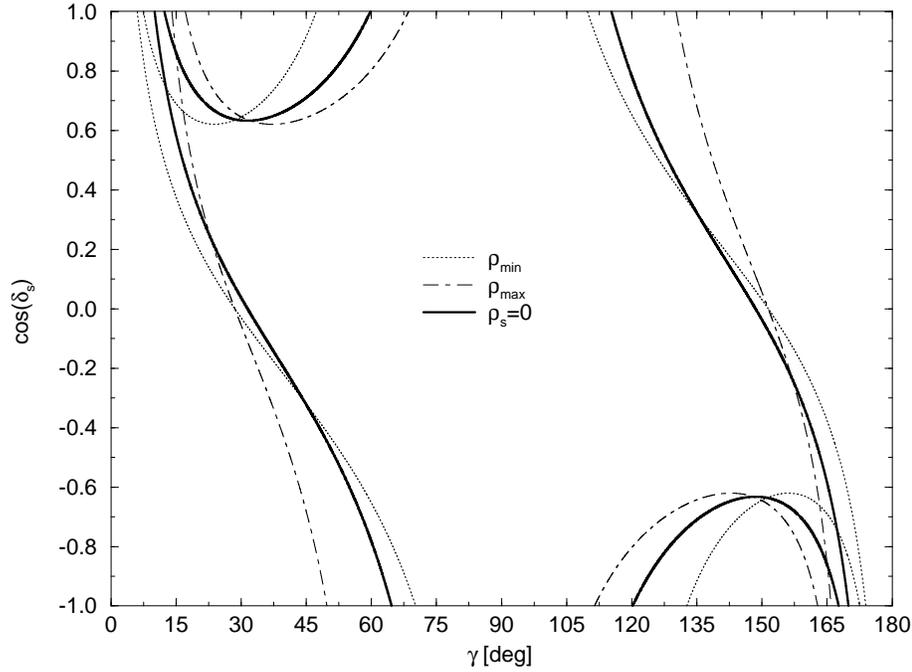}}}
\caption{The contours in the $\gamma\,$--$\,\cos\delta_s$ plane corresponding 
to $(\rho_s)^{\rm max}_{\rm min}$ for $R_H/R_L=0.025$ and 
$(a,b)=(0.60,0.15)$, $(1.40,0.15)$.}
\label{fig:FSI-det}
\end{figure}

If it were possible to measure the ratio $R_H/R_L$ of the observables
of the untagged $B_s\to K^0\overline{K^0}$ decay rate, either $\theta_s$
or $\rho_s$ could be eliminated in (\ref{A-Def})--(\ref{C-Def}), and
the quantity $w_s$ defined by (\ref{ws-def}) could be fixed through
\begin{equation}
w_s=\rho_s\,\sqrt{1+\frac{R_L}{R_H}}\,|\sin\gamma|\,.
\end{equation}
The upper and lower bounds for $\rho_s$ that are implied by $R_H/R_L$
also play an important role to constrain the rescattering effects in the 
$\gamma\,$--$\,\cos\delta_s$ plane. If we assume, for illustrative purposes,
that $R_H/R_L=0.025$ has been measured, and insert (\ref{rhos-bounds}) and 
(\ref{thetas-bounds}) in (\ref{A-Def})--(\ref{C-Def}), we obtain the 
curves shown in Fig.~\ref{fig:FSI-det}, demonstrating nicely the way in 
which the final-state interaction effects can be controlled. By the time 
the $B_s\to K\overline{K}$ observables can be measured, we will probably 
have deeper insights into rescattering processes from analyses of
$B^\pm\to K^\pm K$ and $B^\pm\to\pi^\pm K$ modes anyway, which -- making use 
of the strategies proposed in \cite{defan,my-FSI} -- may provide stringent
constraints on the parameter~$\rho$.

\subsection{The Role of Electroweak Penguins}
\begin{figure}
\centerline{
\rotate[r]{
\epsfxsize=9.2truecm
\epsffile{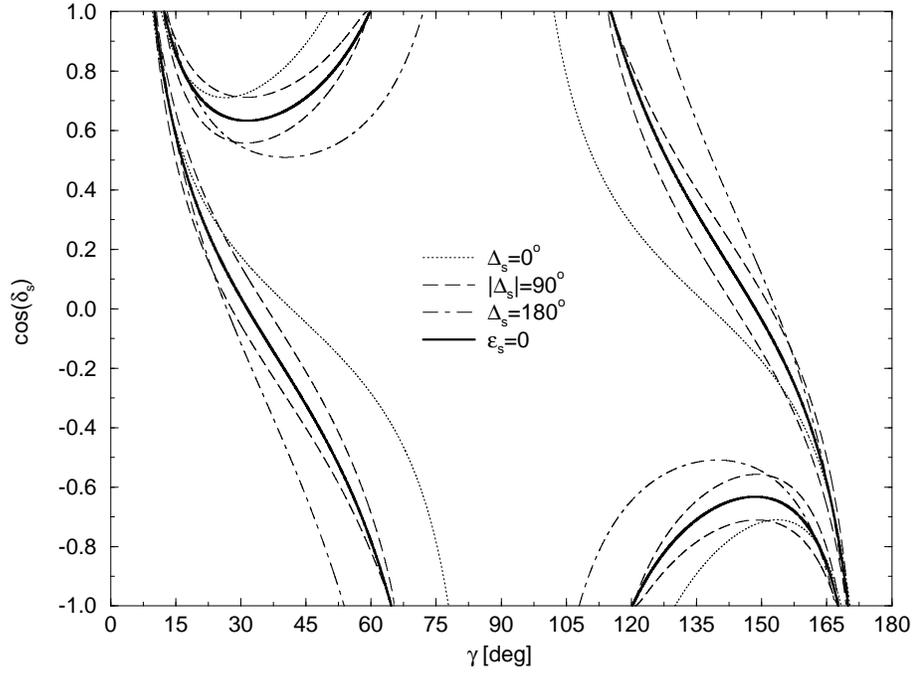}}}
\caption{The contours in the $\gamma\,$--$\,\cos\delta_s$ plane corresponding
to $(a,b)=(0.60,0.15)$ and $(1.40,0.15)$ in the presence of electroweak 
penguins described by $\epsilon_s=0.1$.}
\label{fig:EW-eps01}
\end{figure}

In order to investigate the impact of electroweak penguins on the constraints
on the CKM angle $\gamma$ arising from untagged $B_s\to K\overline{K}$
decays, let us neglect the rescattering effects discussed in the previous
subsection. Combining (\ref{a-def}) and (\ref{b-def}), we obtain 
\begin{equation}
H_\epsilon\,\cos\delta_s\,+\,K_\epsilon\,\sin\delta_s=M_\epsilon
\end{equation}
with
\begin{equation}\label{HK-def}
H_\epsilon\,=\,1\,+\,\epsilon_s\,\cos\Delta_s\,,\quad 
K_\epsilon\,=\,\epsilon_s\,\sin\Delta_s\,,
\end{equation}
\begin{equation}\label{M-def}
M_\epsilon\,=\,\frac{|\sin\gamma|}{2\,\sqrt{b}\,\cos\gamma}
\left[1\,+\,2\,\epsilon_s\cos\Delta_s\,+\,\epsilon_s^2\,-\,a\,+\,b
\left(\frac{\cos\gamma}{\sin\gamma}\right)^2\right],
\end{equation}
yielding
\begin{equation}
\cos\delta_s=\frac{H_\epsilon\,M_\epsilon\,\pm\,K_\epsilon\,
\sqrt{H_\epsilon^2+K_\epsilon^2-M_\epsilon^2}}{H_\epsilon^2+K_\epsilon^2}\,.
\end{equation}
The corresponding contours in the $\gamma\,$--$\,\cos\delta_s$ plane 
are illustrated in Fig.~\ref{fig:EW-eps01} for the $B_s\to K^+K^-$
observables $(a,b)=(0.60,0.15)$ and $(1.40,0.15)$, $\epsilon_s=0.1$, and
various values of the strong phase difference $\Delta_s$. These curves are 
similar to those shown in Fig.~\ref{fig:FSI-rho015} describing the final-state
interaction effects. The electroweak penguin effects are negligibly small for 
$\Delta_s=\pm\,90^\circ$, and maximal for $\Delta_s\in\{0^\circ,180^\circ\}$, 
leading to an uncertainty of $\Delta\gamma=\pm\,11^\circ$ in this example.  

\begin{figure}
\centerline{
\rotate[r]{
\epsfxsize=9.2truecm
\epsffile{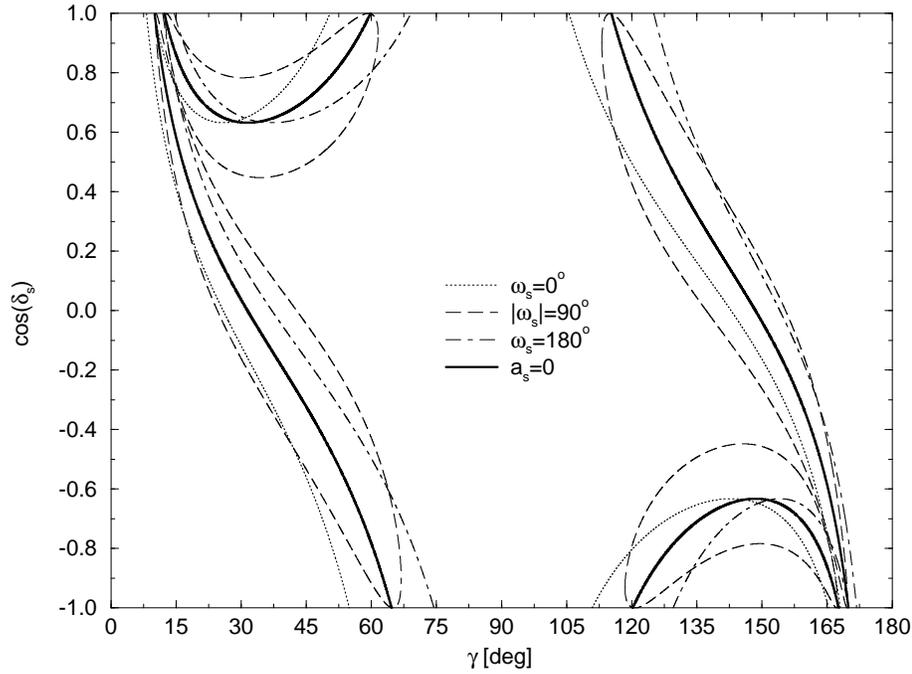}}}
\caption{The contours in the $\gamma\,$--$\,\cos\delta_s$ plane corresponding
to $(a,b)=(0.60,0.15)$ and $(1.40,0.15)$ in the presence of electroweak 
penguins described by $a_s=0.25$.}
\label{fig:EW-a025}
\end{figure}

In the case of $B_{u,d}\to\pi K$ and $B_s\to K\overline{K}$ decays, 
electroweak penguins contribute only in ``colour-suppressed'' form; estimates 
based on simple calculations performed at the perturbative quark level, 
where the relevant hadronic matrix elements are treated within 
the ``factorization'' approach, typically give $\epsilon_{(s)}={\cal O}(1\%)$ 
\cite{fm2}. Since these crude estimates may, however, well underestimate the 
role of electroweak penguins \cite{groro,neubert,fm3}, we have used 
$\epsilon_s=0.1$ in Fig.~\ref{fig:EW-eps01}. Consequently, an improved 
theoretical description of electroweak penguins is highly desirable. 
In Ref.~\cite{defan}, the expressions of the electroweak penguin operators 
and the isospin symmetry of strong interactions have been used to derive the 
following expression:
\begin{equation}\label{eps-r-final}
\frac{\epsilon}{r}\,e^{i(\Delta-\delta)}\approx 0.75
\times a_{u,d}\,e^{i\omega_{u,d}}\,,
\end{equation}
where 
\begin{equation}
a_{u,d}\,e^{i\omega_{u,d}}\equiv\frac{a_2^{\rm eff}}{a_1^{\rm eff}}
\end{equation}
is the ratio of generalized, complex colour factors corresponding to the
usual coefficients $a_1$ and $a_2$. A similar relation holds also for
the $B_s\to K\overline{K}$ parameters. Consequently, we have
\begin{equation}\label{eps-r-rel}
\epsilon_s=q_s\,r_s\,,\quad \Delta_s=\delta_s+\omega_s
\end{equation}
with $q_s\approx0.75\times a_s$. Using (\ref{eps-r-rel}), the definitions
(\ref{HK-def}) and (\ref{M-def}) are modified as follows:
\begin{equation}\label{HK-mod-def}
H_\epsilon\,\to\,H_q\,=\,\cos\gamma\,-\,q_s\,\cos\omega_s\,,\quad K_\epsilon\,
\to\,K_q\,=\,q_s\,\sin\omega_s\,,
\end{equation}
\begin{equation}\label{M-mod-def}
M_\epsilon\,\to\,M_q\,=\,\frac{|\sin\gamma|}{2\,\sqrt{b}}
\left[1\,-\,a\,+\,\frac{b}{\sin^2\gamma}\left(\cos^2\gamma\,-\,2\,q_s
\cos\omega_s\,\cos\gamma\,+\,q_s^2\right)\right].
\end{equation}
In Fig.~\ref{fig:EW-a025}, we show the corresponding contours in the 
$\gamma\,$--$\,\cos\delta_s$ plane for the same observables $a$ and $b$
as in our previous examples, $a_s=0.25$, and various values of the
CP-conserving strong phase $\omega_s$. A first step towards constraining
the electroweak penguin contributions experimentally is provided by the mode
$B^+\to\pi^+\pi^0$, and is closely related to the issue of 
``colour-suppression'', as discussed in detail in \cite{defan}.

\subsection{Combined Rescattering and Electroweak Penguin Effects}
The formulae describing combined rescattering and electroweak penguin 
effects in the contours in the $\gamma\,$--$\,\cos\delta_s$ plane are 
rather complicated. Eliminating $r_s$ in (\ref{a-def}) through (\ref{rs-det}) 
yields an equation to determine $\cos\delta_s$, which has a form similar to
(\ref{cent-eq}):
\begin{equation}
\tilde A\,\cos^2\delta_s\,+\,2\,\tilde B\,\cos\delta_s\,\sin\delta_s\,-\,
\tilde C\,=\,0,
\end{equation}
and the solution 
\begin{equation}
\cos^2\delta_s=\frac{\tilde A\,\tilde C\,+\,2\,\tilde B^2\,\pm\,2\,
\tilde B\,\sqrt{\tilde A\,\tilde C\,+\,\tilde B^2\,-\,\tilde C^2}}{\tilde 
A^2\,+\,4\,\tilde B^2}\,.
\end{equation}
The quantities $\tilde A$, $\tilde B$ and $\tilde C$ are 
given by
\begin{eqnarray}
\tilde A&=&\left(\tilde h^2\,-\,\tilde k^2\right)\tilde D\,+\,4\,\tilde E\,
\frac{\rho_s}{w_s^2}\left(\tilde h\,\cos\theta_s\,-\,
\tilde k\,\sin\theta_s\right)\cos\gamma\\
\tilde B&=&\tilde h\,\tilde k\,\tilde D\,+\,2\,\tilde E\,\frac{\rho_s}{w_s^2}
\left(\tilde k\,\cos\theta_s\,+\,\tilde h\,\sin\theta_s\right)\cos\gamma\\
\tilde C&=&\tilde E^2\,-\,4\,\tilde E\,\frac{\rho_s}{w_s^2}\,\tilde k\,
\sin\theta_s\,\cos\gamma\,-\,\tilde k^2\tilde D
\end{eqnarray}
with
\begin{equation}
\tilde h=1\,+\,\epsilon_s\, w_s\,\cos\Delta_s\,,\quad 
\tilde k=\epsilon_s\, w_s\,\sin\Delta_s
\end{equation}
and
\begin{eqnarray}
\tilde D&=&\frac{4}{w^2_s}\left[\left(\frac{b}{\sin^2\gamma}\right)\,-\,
\frac{\rho_s^2}{w_s^2}\right]\cos^2\gamma\\
\tilde E&=&1-a+b\left(\frac{\cos\gamma}{\sin\gamma}\right)^2-\,
\frac{\rho_s^2}{w_s^2}+2\,\frac{\epsilon_s}{w_s}\,\biggl[\cos\Delta_s+
\rho_s\,\cos(\Delta_s-\theta_s)\,\cos\gamma\biggr]+\,\epsilon_s^2\,.
\end{eqnarray}
In order to derive these expressions, no approximations have been made and
they are valid exactly.

\boldmath
\section{Combining $B_{u,d}\to\pi K$ and Untagged $B_s\to K\overline{K}$ 
Decays Through $SU(3)$ Flavour Symmetry}\label{SU3}
\unboldmath
So far, we have discussed the $B_{u,d}\to\pi K$ and $B_s\to K\overline{K}$
decays separately. As we have just seen, interesting constraints on 
$\gamma$ may arise from the corresponding observables. The goal is, 
however, not only to constrain, but eventually to determine $\gamma$.
If we consider the modes $B_{u,d}\to\pi K$ and 
$B_s\to K\overline{K}$ separately, information about the magnitudes 
of the amplitudes $T$ and $T_s$ is needed to accomplish this task 
\cite{PAPIII,groro,fd1}. Such an input can be avoided, if the $SU(3)$ 
flavour symmetry of strong interactions is applied, yielding
\begin{equation}\label{SU3-input}
\cos\delta_s=\zeta_\delta\,\cos\delta\,,\quad r_s=\zeta_r\, r\,, 
\end{equation}
where the quantities $\zeta_\delta$ and $\zeta_r$ parametrize $SU(3)$-breaking
corrections. As a first ``guess'', we may use $\zeta_\delta=1$, $\zeta_r=1$, 
which corresponds to the strict $SU(3)$ limit. In order to extract $\gamma$ 
from a simultaneous analysis of $B_{u,d}\to\pi K$ and $B_s\to K\overline{K}$ 
decays, each of the two expressions given in (\ref{SU3-input}) is in principle 
sufficient. The former applies to the contours in the 
$\gamma\,$--$\,\cos\delta_{(s)}$ plane, while the latter can be used for the 
$\gamma\,$--$\,r_{(s)}$ plane. 

\begin{figure}
\centerline{
\rotate[r]{
\epsfxsize=9.2truecm
\epsffile{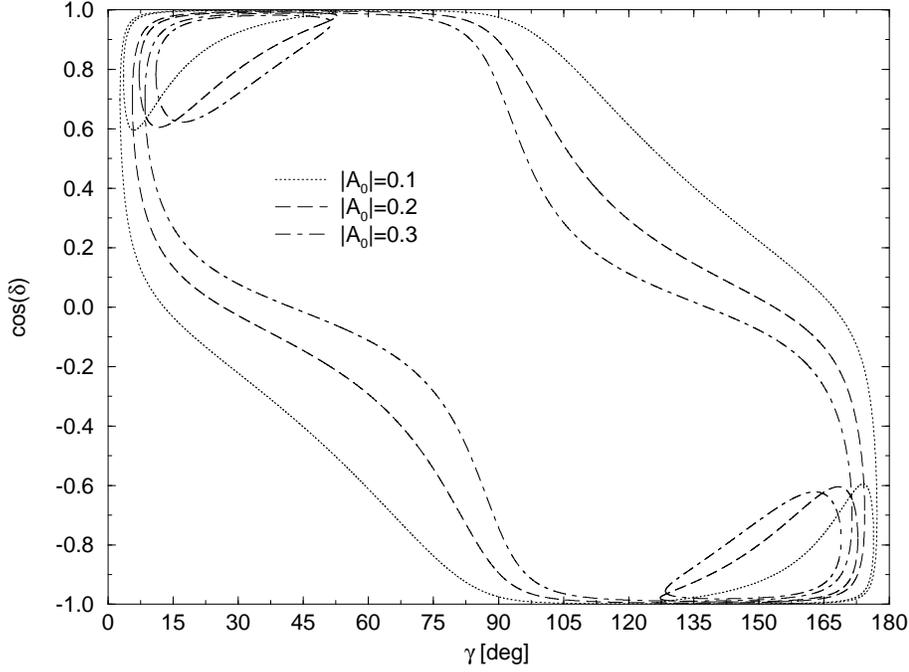}}}
\caption{The contours in the $\gamma\,$--$\,\cos\delta$ plane for 
$R=0.65$, $1.05$ and various values of $A_0$ in the case of neglected 
rescattering and electroweak penguin effects.}\label{fig:CD-cont}
\end{figure}

Let us first turn to the contours in the $\gamma\,$--$\,\cos\delta_{(s)}$ 
plane. In the case of the $B_s\to K\overline{K}$ decays, these contours
play a key role to constrain the CKM angle $\gamma$, while it is more
``natural'' to consider contours in the $\gamma\,$--$\,r$ plane in the
case of the $B_{u,d}\to\pi K$ modes \cite{defan}. However, the 
$B_{u,d}\to\pi K$ observables $R$ and $A_0$ fix also contours in the 
$\gamma\,$--$\,\cos\delta$ plane. If we eliminate $r$ in $R$ through 
the ``pseudo-asymmetry'' $A_0$ (see (\ref{R-exp}) and (\ref{A0-exp})),
we obtain
\begin{equation}
\cos^2\delta\,=\,\frac{2\,V^2\,-\,U\,W\,\pm\,2\,V\,\sqrt{V^2\,-\,U\,W\,-\,
W^2}}{U^2\,+\,4\,V^2}
\end{equation}
with
\begin{eqnarray}
U&=&2\,A\left(k\,B\,+\,h\,C\,\right)\,+\,\left(B^2\,-\,C^2\right)
\left(R\,-\,R_0\right)\\
V&=&B\,C\left(R\,-\,R_0\right)\,-\,A\left(h\,B\,-\,k\,C\right)\\
W&=&A^2\,-\,2\,k\,A\,B\,-\,\left(R\,-\,R_0\right)B^2\,,
\end{eqnarray}
where the quantities 
\begin{equation}
h=\frac{1}{w}\left(\cos\gamma+\rho\,\cos\theta\right)+\epsilon\,\cos\Delta
\,\cos\gamma\,,
\quad k=\frac{\rho}{w}\,\sin\theta+\epsilon\,\sin\Delta\,\cos\gamma
\end{equation}
\begin{equation}
A=\frac{A_0-A_+}{2\,\sin\gamma}-\frac{\epsilon\,\rho}{w}\,
\sin(\Delta-\theta)\,,\quad
B=\frac{1}{w}+\epsilon\,\cos\Delta\,,\quad
C=\epsilon\,\sin\Delta\,,
\end{equation}
and 
\begin{equation}
R_0=1+2\,\frac{\epsilon}{w}\,\left[\,\cos\Delta+\rho\,\cos(\Delta-\theta)
\cos\gamma\,\right]+\epsilon^2
\end{equation}
were introduced in \cite{defan}. The corresponding contours in the 
$\gamma\,$--$\,\cos\delta$ plane are illustrated in Fig.~\ref{fig:CD-cont}
for $R=0.65$, $1.05$, and various values of $A_0$ in the case of 
neglected rescattering and electroweak penguin effects. Using the formulae
given above and the strategies proposed in \cite{defan,my-FSI}, these effects 
can be included in these contours. For $R=0.65$, a significant range around 
$\gamma=90^\circ$ is excluded, which corresponds to $R<R_{\rm min}$, where 
$R_{\rm min}$ is given in (\ref{Rmin}). If the contours arising in the 
$\gamma\,$--$\,\cos\delta_s$ plane determined from the untagged 
$B_s\to K\overline{K}$ decays are included in the same figure (see 
Fig.~\ref{fig:CD0}), $\gamma$ and $\cos\delta_{(s)}$ can be determined with 
the help of the first relation given in (\ref{SU3-input}). 

Another approach to accomplish this task is to consider the contours in 
the $\gamma\,$--$\,r_{(s)}$ plane. To this end, we rewrite the observable
$a$ given in (\ref{a-def}) as
\begin{equation}
a=a_0\,-\,2\,r_s\left(f\,\cos\delta_s\,+\,g\,\sin\delta_s\right)\,+
\,r_s^2\cos^2\gamma
\end{equation}
with
\begin{equation}
a_0=1+2\,\frac{\epsilon_s}{w_s}\left[\cos\Delta_s+\rho_s\,\cos(\Delta_s-
\theta_s)\,\cos\gamma\right]+\epsilon_s^2-\frac{\rho_s^2}{w_s^2}\,
\sin^2\gamma
\end{equation}
\begin{equation}
f=\left[\frac{1}{w_s}\left(1+\rho_s\cos\theta_s\cos\gamma\right)+
\epsilon_s\cos\Delta_s\right]\cos\gamma,\,\,\, g=\left(
\frac{\rho_s}{w_s}\sin\theta_s\cos\gamma+\epsilon_s\sin\Delta_s\right)
\cos\gamma,
\end{equation}
and eliminate the strong phase $\delta_s$ in (\ref{b-def}), yielding
\begin{equation}
r_s=\sqrt{S\,\pm\,\sqrt{S^2\,-\,T}}\,,
\end{equation}
where
\begin{equation}
S=\frac{p\,q+l^2\,m}{q^2+l^2\cos^4\gamma}\,,\quad T=\frac{p^2+
(a-a_0)^2\,l^2}{q^2+l^2\cos^4\gamma}
\end{equation}
with 
\begin{eqnarray}
p&=&\frac{b}{\sin^2\gamma}-\frac{\rho_s^2}{w_s^2}-\frac{\rho_s}{w_s}\,
(a-a_0)\left(\frac{f\cos\theta_s+g\,\sin\theta_s}{f^2+g^2}\right)\\
q&=&1-\frac{\rho_s}{w_s}\left(\frac{f\cos\theta_s+g\,\sin\theta_s}{f^2+g^2}
\right)\cos^2\gamma\\
l&=&\frac{\rho_s}{w_s}\left(\frac{g\,\cos\theta_s-
f\sin\theta_s}{f^2+g^2}\right)\\
m&=&2\,(f^2+g^2)+(a-a_0)\cos^2\gamma\,.
\end{eqnarray}

\begin{figure}
\centerline{
\rotate[r]{
\epsfxsize=9.2truecm
\epsffile{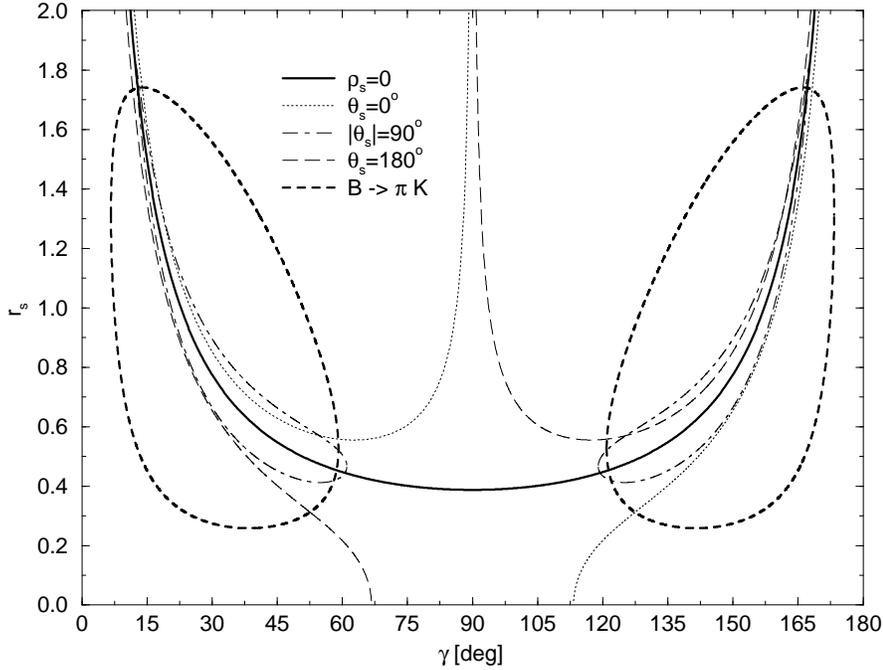}}}
\caption{The contours in the $\gamma\,$--$\,r_s$ plane for 
$a=0.60$, $b=0.15$ and $\rho_s=0.15$, $\epsilon_s=0$. The 
$B_{u,d}\to\pi K$ contours correspond to $R=0.75$, $A_0=0.2$ and 
$\rho=\epsilon=0$.}\label{fig:r-cont}
\end{figure}

In Fig.~\ref{fig:r-cont}, we have illustrated the corresponding contours
in the $\gamma\,$--$\,r_s$ plane by choosing $a=0.60$ and $b=0.15$. The thick
solid line has been calculated for $\rho_s=0$, while the thin lines have 
been obtained for $\rho_s=0.15$ and various values of the strong phase 
$\theta_s$. The effects of electroweak penguins have been neglected in
this figure. Since the observable $b$ is not affected by electroweak 
penguins at all, as can be seen in (\ref{b-def}), their effect on the 
contours in the $\gamma\,$--$\,r_s$ plane is rather small and is only due to 
the elimination of $\delta_s$ through the observable $a$. In 
Fig.~\ref{fig:r-cont}, we have also included the contours arising from the 
$B_{u,d}\to\pi K$ observables $R=0.75$ and $A_0=0.2$ in the case of 
$\zeta_r=1$ and neglected rescattering and electroweak penguin effects, 
which can be taken into account by using the strategies presented in 
\cite{defan,my-FSI}. Combining the $B_s\to K\overline{K}$ and 
$B_{u,d}\to\pi K$ contours, $\gamma$ and $r_{(s)}$ can be determined, 
as illustrated in Fig.~\ref{fig:r-cont}. The $SU(3)$-breaking parameter 
$\zeta_r$ can be expressed as
\begin{equation}
\zeta_r=\sqrt{\frac{\left\langle|P|^2\right\rangle}{\left\langle|P_s|^2
\right\rangle}}\frac{|T_s|}{|T|}\,,
\end{equation}
where $\left\langle|P|^2\right\rangle$ and $\left\langle|P_s|^2\right\rangle$
can be fixed experimentally through the combined $B^\pm\to\pi^\pm K$ 
branching ratio (\ref{BR-char}) and the untagged $B_s\to K^0\overline{K^0}$ 
rate (\ref{Psaver}), respectively. Comparing their values with those of
$R$ and $R_s$, we obtain interesting insights into $SU(3)$ breaking. The 
amplitudes $T$ and $T_s$ are given in (\ref{T-def}) and (\ref{Ts-def}),
respectively, and their structure is quite similar. Note that the different
signs of the ${\cal A}$ and $\tilde{\cal E}$ contributions are only due to 
our definition of meson states (see, for instance, \cite{ghlr}). In the 
strict $SU(3)$ limit, the combinations $\tilde{\cal T}-{\cal A}$ and 
$\tilde{\cal T}_s+\tilde{\cal E}_s$ would be equal. 

Assuming $|T|\approx |T_s|$ is probably more reliable than 
$\cos\delta_s\approx\cos\delta$. However, as soon as the $B_{u,d}\to\pi K$ 
observables $R$ and $A_0$, as well as the $B_s\to K\overline{K}$ 
observables $a$ and $b$ have been measured, both the contours in the 
$\gamma\,$--$\,r_{(s)}$ plane and those in the 
$\gamma\,$--$\,\cos\delta_{(s)}$ plane should be considered to extract the 
CKM angle $\gamma$ and the hadronic parameters $r_{(s)}$ and 
$\cos\delta_{(s)}$ as discussed above. In addition to $\gamma$, in particular 
$r_{(s)}$ would be of special interest, since it allows a test of the 
factorization hypothesis by comparing its experimentally determined value 
with (\ref{r-fact}).

\begin{figure}
\centerline{
\rotate[r]{
\epsfxsize=7.5truecm
\epsffile{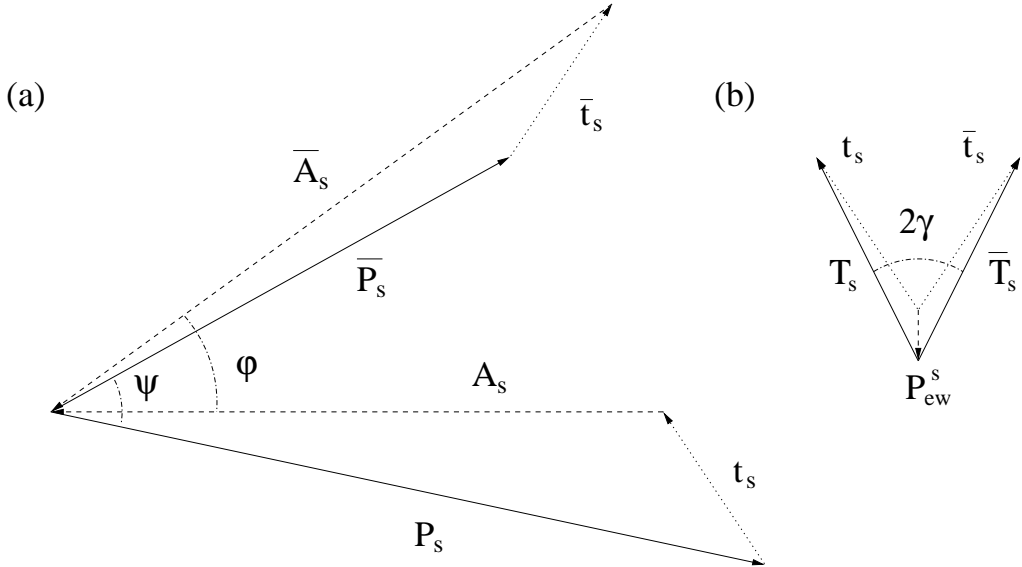}}}
\vspace*{0.7truecm} 
\caption{Representation of the amplitude relations (\ref{ampl-neuts}) 
and (\ref{ampl-chars}) between the decays $B_s^0\to K^+K^-$, $B_s^0\to 
K^0\overline{K^0}$ and their charge conjugates in the complex 
plane.}\label{fig:triangles}
\end{figure}

\boldmath
\section{A Brief Look at a Tagged Analysis of 
$B_s\to K\overline{K}$}\label{tagged}
\unboldmath
In this paper, we have so far focused on the observables provided by an 
{\it untagged} measurement of the decays $B_s\to K^0\overline{K^0}$ and 
$B_s\to K^+K^-$, where the rapid oscillatory $\Delta M_st$-terms cancel. 
Let us briefly discuss, in this section, an interesting feature of a 
{\it tagged} analysis of these modes. If the $\Delta M_st$-oscillations 
could be resolved in such a measurement, the CKM angle $\gamma$ could be 
extracted by using only the amplitude relations (\ref{ampl-neuts}) and 
(\ref{ampl-chars}), which are based on the $SU(2)$ isospin symmetry of 
strong interactions and are completely general. These relations can be 
represented in the complex plane as triangles for the $B_s^0\to K\overline{K}$ 
decays and their charge conjugates, as shown in Fig.~\ref{fig:triangles}, 
where $A_s\equiv A(B^0_s\to K^+K^-)$, 
$P_s\equiv A(B^0_s\to K^0\overline{K^0})$, and $t_s\equiv T_s+P_{\rm ew}^s$. 
The angles $\varphi$ and $\psi$ can be determined by measuring the 
time-dependent CP asymmetries arising in $B^0_s\to K^+K^-$ and 
$B^0_s\to K^0\overline{K^0}$. These CP-violating asymmetries are defined 
for a generic final state $f$ by
\begin{equation}
a_{\rm CP}(t;f)\equiv\frac{\Gamma(B_s^0(t)\to f)\,-\,
\Gamma(\overline{B^0_s}(t)\to f)}{\Gamma(B_s^0(t)\to f)\,+\,
\Gamma(\overline{B^0_s}(t)\to f)}\,.
\end{equation}
If $f$ is a CP eigenstate, we have
\begin{equation}
a_{\rm CP}(t;f)=2\,e^{-\Gamma_st}\left[\frac{{\cal A}_{\rm CP}^{\rm
dir}(B_s\to f)\cos(\Delta M_st)+{\cal A}_{\rm CP}^{\rm mix-ind}(B_s\to f)
\sin(\Delta M_st)}{e^{-\Gamma_H^{(s)}t}+e^{-\Gamma_L^{(s)}t}\,+\,
{\cal A}_{\Delta\Gamma}(B_s\to f)
\left(e^{-\Gamma_H^{(s)}t}\,-\,e^{-\Gamma_L^{(s)}t}\right)}\right],
\end{equation}
where
\begin{equation}
{\cal A}_{\rm CP}^{\rm dir}(B_s\to f)=\frac{1-\left|\xi_f\right|^2}{1+
\left|\xi_f\right|^2}\,,\quad {\cal A}_{\rm CP}^{\rm mix-ind}(B_s\to f)=
\frac{2\,\mbox{Im}(\xi_f)}{1+\left|\xi_f\right|^2}\,,
\end{equation}
and 
\begin{equation}
{\cal A}_{\Delta\Gamma}(B_s\to f)=
\frac{2\,\mbox{Re}(\xi_f)}{1+\left|\xi_f\right|^2}
\end{equation}
can be determined straightforwardly from the untagged rate given in 
(\ref{Bsrate}). The observable $\xi_f$ has been introduced in (\ref{xi-def}). 
In the case of the decays $B_s\to K^+K^-$ and $B_s\to K^0\overline{K^0}$, 
we have
\begin{eqnarray}
{\cal A}_{\rm CP}^{\rm dir}(B_s\to K^+K^-)&=&
\frac{|A_s|^2-|\overline{A_s}|^2}{|A_s|^2+|\overline{A_s}|^2}\\
{\cal A}_{\rm CP}^{\rm mix-ind}(B_s\to K^+K^-)&=&
\frac{2\,|A_s||\overline{A_s}|}{|A_s|^2+|\overline{A_s}|^2}\,\sin
\left(\phi_{\rm M}^{(s)}+\varphi\right)
\end{eqnarray}
and
\begin{eqnarray}
{\cal A}_{\rm CP}^{\rm dir}(B_s\to K^0\overline{K^0})&=&
\frac{|P_s|^2-|\overline{P_s}|^2}{|P_s|^2+|\overline{P_s}|^2}\\
{\cal A}_{\rm CP}^{\rm mix-ind}(B_s\to K^0\overline{K^0})&=&
\frac{2\,|P_s||\overline{P_s}|}{|P_s|^2+|\overline{P_s}|^2}\,\sin
\left(\phi_{\rm M}^{(s)}+\psi\right),
\end{eqnarray}
where $\phi_{\rm M}^{(s)}=0$ to a good approximation in the Standard Model. 
The observables ${\cal A}_{\Delta\Gamma}(B_s\to K^+K^-)$ and 
${\cal A}_{\Delta\Gamma}(B_s\to K^0\overline{K^0})$ probe $\cos\left(
\phi_{\rm M}^{(s)}+\varphi\right)$ and $\cos\left(\phi_{\rm M}^{(s)}+\psi
\right)$, respectively. Consequently, the CP-violating observables allow the 
construction of the amplitudes $A_s$, $\overline{A_s}$ and $P_s$, 
$\overline{P_s}$ in the complex plane, i.e.\ of the dashed and solid lines 
in Fig.~\ref{fig:triangles} (a). So far, we have not made any approximations, 
and this construction is valid exactly. However, in order to extract 
$\gamma$, we have to neglect the electroweak penguin amplitude 
$P_{\rm ew}^s$. Then we have $t_s=T_s$, and 
the relative orientation of the $A_s$, $P_s$, $T_s$ and $\overline{A_s}$, 
$\overline{P_s}$, $\overline{T_s}$ amplitudes can be fixed by requiring 
$|T_s|=|\overline{T_s}|$. The angle between $T_s$ and $\overline{T_s}$ is 
given by $2\gamma$. In Fig.~\ref{fig:triangles} (b), the impact of 
electroweak penguins on this construction is illustrated for 
$\omega_s=0^\circ$ (see (\ref{eps-r-rel})). 

From a conceptual point of view, this determination of $\gamma$ is quite
analogous to that of the extraction of the CKM angle $\alpha$ from a 
combined analysis of the decays $B_d\to\pi^+\pi^-$ and $B_d\to K^0
\overline{K^0}$, which was proposed in \cite{PAPII}. Recently, a study
of tagged $B_s\to K^0\overline{K^0}$ and $B_s\to K^+K^-$ decays was 
performed in \cite{kly}, where also the possibility of extracting $\gamma$ 
from these decays was pointed out. However, in that paper rescattering 
and electroweak penguin effects were neglected. Here we have shown that
a time-dependent analysis of tagged $B_s\to K\overline{K}$ decays
allows a determination of $\gamma$ taking into account rescattering
effects ``automatically''. This approach works also if new physics 
contributes to $B^0_s$--$\overline{B^0_s}$ mixing.

\section{Physics Beyond the Standard Model}\label{BSM}
Let us focus in this section on a particular scenario of new physics 
\cite{new-phys,HF97-talk}, where the $B_{u,d}\to\pi K$ and 
$B_s\to K\overline{K}$ modes are governed by the Standard Model diagrams and 
$B^0_s$--$\overline{B^0_s}$ mixing receives significant CP-violating 
new-physics contributions. A similar scenario for the $B_{u,d}\to\pi K$ 
modes and $B^0_d$--$\overline{B^0_d}$ mixing was considered in \cite{gnps}.
While the CP-violating weak mixing phase of the $B_s$ system vanishes to 
a good approximation within the Standard Model, as we have already noted, 
new physics may lead to a sizeable CP-violating weak 
$B^0_s$--$\overline{B^0_s}$ mixing phase. Applying the same notation as 
in \cite{rev,HF97-talk}, we have
\begin{equation}
\phi_{\rm M}^{(s)}=0+2\,\phi_{\rm new}^{(s)}\equiv2\,\phi\,.
\end{equation}
If the decay $B_s\to K^0\overline{K^0}$ is still dominated by the Standard 
Model contributions, the observables $R_L$ and $R_H$ of the corresponding 
untagged transition rate introduced in (\ref{RL}) and (\ref{RH}) are modified 
as follows: 
\begin{equation}\label{RL-NP}
R_L=\left[\cos^2\phi+2\,\rho_s\cos\theta_s\cos
\phi\cos\left(\phi+\gamma\right)+
\rho_s^2\cos^2\left(\phi+\gamma\right)\right]\Gamma_0
\end{equation}
\begin{equation}\label{RH-NP}
R_H=\left[\,\sin^2\phi\,+\,2\,\rho_s\cos\theta_s
\sin\phi\sin\left(\phi+\gamma\right)+
\rho_s^2\sin^2\left(\phi+\gamma\right)
\,\right]\Gamma_0\,.
\end{equation}
Note that the new-physics contributions to $R_L$ and $R_H$ cancel in their
sum, taking the same form as (\ref{Psaver}).  

\begin{figure}
\centerline{
\rotate[r]{
\epsfxsize=9.2truecm
\epsffile{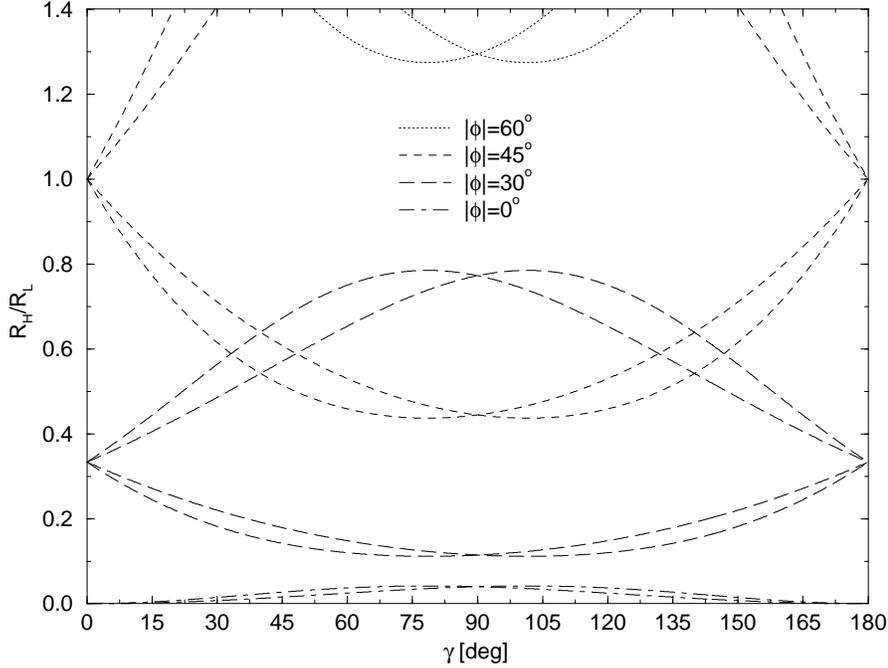}}}
\caption{The dependence of $R_H/R_L$ on the CKM angle $\gamma$ in the
presence of CP-violating new-physics contributions to 
$B^0_s$--$\overline{B^0_s}$ mixing for $\rho_s=0.2$ and $\theta_s
\in\{0^\circ,180^\circ\}$.}\label{fig:RHRLNP}
\end{figure}

While rescattering effects may lead to values of $R_H/R_L$ of at most
${\cal O}(10\%)$ in the Standard Model, as we have seen in 
Subsection~\ref{res-effects}, in the scenario of new physics considered
here, this ratio may be enhanced dramatically. This feature is shown in
Fig.~\ref{fig:RHRLNP} for various values of the mixing phase 
$\phi$ and $\rho_s=0.2$. The upper and lower curves shown for each 
value of $\phi$ correspond to $\theta_s=0^\circ$ and
$180^\circ$. Note that $\rho_s$ enters in (\ref{RH-NP}) already at
the ${\cal O}(\rho_s)$ level, in contrast to (\ref{RH}). 

The CP-violating new-physics phase $\phi$ can be extracted, for
instance, from the decays $B_s\to J/\psi\,\phi$ or $B_s\to D_s^+ D_s^-$
\cite{new-phys,HF97-talk}. In contrast to $B_s\to K^0\overline{K^0}$, which is 
a ``rare'' penguin-induced decay, these channels are governed by tree-level 
processes. Consequently, it is plausible to assume that their decay amplitudes 
are in general affected to a smaller extent by physics beyond the Standard 
Model than that of $B_s\to K^0\overline{K^0}$ \cite{gw}. Since an analysis of
$B_s\to J/\psi(\to l^+l^-)\,\phi(\to K^+K^-)$ requires consideration of the
angular distributions of its decay products \cite{fd1,ddf1}, let us here
focus on the mode $B_s\to D_s^+ D_s^-$. Its untagged decay rate is given 
by \cite{fd1}
\begin{equation}\label{D+D-NP}
\Gamma[D_s^+D_s^-(t)]\equiv D_L\,e^{-\Gamma_L^{(s)}t}+ D_H\,
e^{-\Gamma_H^{(s)}t}=\left[\,\cos^2\phi\,e^{-\Gamma_L^{(s)}t}+
\sin^2\phi\,e^{-\Gamma_H^{(s)}t}\,\right]\tilde\Gamma_0
\end{equation}
and allows the determination of
\begin{equation}\label{phiNP-det}
\cos2\,\phi=\frac{D_L-D_H}{D_L+D_H}\equiv{\cal D}\,.
\end{equation}
The corresponding combination of the $B_s\to K^0\overline{K^0}$ observables
$R_L$ and $R_H$ takes on the other hand the form
\begin{eqnarray}
{\cal R}\equiv\frac{R_L-R_H}{R_L+R_H}&=&\left[\,1-
\frac{2\,\rho_s^2\sin^2\gamma}{1+2\,\rho_s\,\cos\theta_s\,\cos\gamma+
\rho_s^2}\,\right]\cos2\,\phi\nonumber\\
&&-\,2\,\rho_s\left[\,\frac{\cos\theta_s+\rho_s\cos\gamma}{1+
2\,\rho_s\,\cos\theta_s\,\cos\gamma+\rho_s^2}\,\right]\sin\gamma\,
\sin2\,\phi\,,
\end{eqnarray}
so that we have
\begin{equation}\label{obs-comb}
\frac{{\cal D}-{\cal R}}{2\,{\cal D}}=\rho_s\,\sin\gamma
\left[\,\frac{\pm\left(\cos\theta_s+\rho_s\cos\gamma\right)
\sqrt{1/{\cal D}^2-1}+\rho_s\sin\gamma}{1+
2\,\rho_s\,\cos\theta_s\,\cos\gamma+\rho_s^2}\,\right]=\,{\cal O}(\rho_s).
\end{equation}
In the case of $|\theta_s|\approx90^\circ$, this quantity is even of 
${\cal O}(\rho_s^2)$. A future measurement of (\ref{obs-comb}) that is 
significantly larger than ${\cal O}(10\%)$ would indicate 
new-physics contributions to the $B_s\to K^0\overline{K^0}$ decay amplitude
(see \cite{fknp,fm3} for discussions of such scenarios). 

Using (\ref{phiNP-det}) to determine $\cos2\,\phi$, we 
encounter a sign ambiguity, since it cannot be decided experimentally 
which one of the two decay widths corresponds to $\Gamma_L^{(s)}$ and 
$\Gamma_H^{(s)}$. While one expects $\Gamma_L^{(s)}>\Gamma_H^{(s)}$ within 
the Standard Model, that need not be the case in the scenario of new physics 
discussed here, leading to the following modification of the Standard Model 
width difference $\Delta\Gamma_s^{\rm SM}<0$ \cite{grossman}:
\begin{equation}\label{Dgamma-NP}
\Delta\Gamma_s=\Delta\Gamma_s^{\rm SM}\cos2\,\phi.
\end{equation}
Consequently, the magnitude of $\Delta\Gamma_s$ is reduced, and for 
$\cos2\,\phi<0$ even its sign is reversed. An interesting
implication of (\ref{phiNP-det}) and (\ref{Dgamma-NP}), which provides a nice
consistency check, is the feature that the larger part $D_{L,H}$ of 
(\ref{D+D-NP}) always enters with a larger decay width $\Gamma^{(s)}_{L,H}$
than the smaller piece $D_{H,L}$.

Assuming, as in (\ref{RL-NP}) and (\ref{RH-NP}), that new physics shows up 
only in $B^0_s$--$\overline{B^0_s}$ mixing, it is a straightforward 
exercise to derive the modified expressions for the observables $a$ and $b$ 
of the untagged $B_s\to K^+K^-$ rate given in (\ref{a-def}) and (\ref{b-def}).
Since the corresponding formulae are rather complicated in the general case, 
and therefore not very instructive, let us just give the expressions for
neglected rescattering and electroweak penguin effects:
\begin{eqnarray}
\left.a\right|_{\rho_s=\epsilon_s=0}&=&\cos^2\phi-2\,r_s\cos\delta_s\cos
\phi\cos\left(\phi+\gamma\right)+
r_s^2\cos^2\left(\phi+\gamma\right)\label{a-NP}\\
\left.b\right|_{\rho_s=\epsilon_s=0}&=&\sin^2\phi\,-\,2\,r_s\cos\delta_s
\sin\phi\sin\left(\phi+\gamma\right)+
r_s^2\sin^2\left(\phi+\gamma\right),\label{b2-NP}
\end{eqnarray}
which are very similar to (\ref{RL-NP}) and (\ref{RH-NP}). In such a scenario 
of physics beyond the Standard Model, the new-physics contributions to $a$ 
and $b$ cancel in their sum $R_s\equiv a+b$. Since the new-physics phase 
$\phi$ can be determined (up to discrete ambiguities) from analyses of 
$B_s\to J/\psi\,\phi$ or $B_s\to D_s^+ D_s^-$ decays, it is 
straightforward to see that the strategies presented in the previous sections 
can still be performed for such a scenario of new physics. In fact, since
$\rho_s$ enters $R_H/R_L$ already at the ${\cal O}(\rho_s)$ level, it
may be considerably easier to constrain $\rho_s$ through this ratio. 
Neglecting terms of ${\cal O}(\rho_s^2)$ yields
\begin{equation}\label{det-y}
\rho_s\cos\theta_s\approx\frac{R_H\cos^2\phi\,-\,
R_L\sin^2\phi}{\left(R_H+R_L\right)
\sin2\phi\,\sin\gamma+2\left(R_L\sin^2
\phi\,-\,R_H\cos^2\phi\right)\cos\gamma}\,.
\end{equation}
A similar comment applies to the observable $b$, receiving only contributions 
from second-order terms ${\cal O}(r_s^2)$, ${\cal O}(r_s\,\rho_s)$ and 
${\cal O}(\rho_s^2)$ in the Standard Model, as can be seen in (\ref{b-def}).
If we neglect all second-order terms of this kind and those involving 
$\epsilon_s$, we obtain
\begin{eqnarray}
a&\approx&\left(\,1\,-\,2\,x\,\cos\gamma\,+\,2\,z\,\right)\cos^2\phi\,+\,
\left(\,x\,-\,y\,\right)\sin\gamma\,\sin2\,\phi\label{approx-a}\\
b&\approx&\left(\,1\,-\,2\,x\,\cos\gamma\,+\,2\,z\,\right)\sin^2\phi\,-\,
\left(\,x\,-\,y\,\right)\sin\gamma\,\sin2\,\phi\,,\label{approx-b}
\end{eqnarray}
where
\begin{equation}
x\equiv r_s\,\cos\delta_s\,,\quad y\equiv\rho_s\,\cos\theta_s\,,\quad z\equiv
\epsilon_s\,\cos\Delta_s\,.
\end{equation}
Since $y$ can be determined with the help of (\ref{det-y}) as a function 
of $\gamma$, it can be eliminated in (\ref{approx-a}) and (\ref{approx-b}). 
Consequently, the $B_s\to K^+K^-$ observables $a$ and $b$ then allow the 
extraction of $\gamma$ and $x$ for a given value of the quantity $z$, 
parametrizing the uncertainty due to electroweak penguins. Let us emphasize 
that {\it untagged} data samples are sufficient to this end, in contrast 
to the strategy discussed in Section~\ref{tagged}, making use of tagged 
$B_s\to K\overline{K}$ decays. Needless to note, another possibility is to 
extract $\gamma$ and $z$ as functions of $x$, or $x$ and $z$ as functions 
of $\gamma$. Neglecting rescattering and electroweak penguin effects, i.e.\ 
$y=z=0$, the corresponding formulae simplify considerably and yield
\begin{equation}
\tan\gamma\approx
\frac{a\,-\,b\,-\,(\,a\,+\,b\,)\,\cos2\,\phi}{(\,1\,-\,a\,-\,b\,)
\,\sin2\,\phi}\,,
\end{equation}
where $\cos2\phi$ is given in (\ref{phiNP-det}) and 
\begin{equation}
|\sin2\phi|=2\,\frac{\sqrt{D_L\,D_H}}{D_L\,+\,D_H}\,. 
\end{equation}
Consequently, within this approximation, the observables of the untagged 
$B_s\to K\overline{K}$ decays provide sufficient information to determine 
the CKM angle $\gamma$. One should keep it in mind, however, that new-physics 
contributions to $B^0_s$--$\overline{B^0_s}$ mixing result also in a 
reduction of the width difference $|\Delta\Gamma_s|$ (see (\ref{Dgamma-NP})), 
which could, on the other hand, make untagged analyses more difficult. 

\section{Conclusions}\label{concl}
In summary, we have seen that $B_{u,d}\to\pi K$ and untagged $B_s\to K
\overline{K}$ decays offer interesting tools to probe the CKM angle $\gamma$.
It is possible to obtain constraints on $\gamma$ both from combined 
$B_d\to\pi^\mp K^\pm$, $B^\pm\to\pi^\pm K$ branching ratios and from the 
time evolution of untagged $B_s\to K^+K^-$, $B_s\to K^0\overline{K^0}$ 
decays. To this end, in the former case the ratio $R$ of combined 
$B_{u,d}\to\pi K$ branching ratios has to be smaller than 1, while in the 
latter case only a sizeable width difference $\Delta\Gamma_s$ is required. 
These bounds on $\gamma$ provide information complementary to the present 
range for this angle arising from the usual fits of the unitarity triangle 
and are hence of particular phenomenological interest. Moreover, also a 
certain range for $\gamma$ around $0^\circ$ and $180^\circ$ can be excluded 
through the untagged $B_s\to K\overline{K}$ decays. In the $B_{u,d}\to\pi K$
case, direct CP violation in $B_d\to\pi^\mp K^\pm$ has to be observed to
accomplish this task.

The theoretical accuracy of these constraints, which make only use of the
general amplitude structure arising within the Standard Model and of the
$SU(2)$ isospin symmetry of strong interactions, is limited by certain
rescattering processes and contributions arising from electroweak penguins. 
In this paper, we have presented a completely general formalism, taking also
into account these effects. The rescattering effects can be controlled in the
case of the $B_{u,d}\to\pi K$ decays by using experimental data on the mode
$B^\pm\to K^\pm K$. Concerning the bounds on $\gamma$, the rescattering
effects can, in this way, be included completely. In the case of the $B_s\to K
\overline{K}$ decays, the observables of the untagged 
$B_s\to K^0\overline{K^0}$ rate play an important role in this respect. 
Moreover, by the time the $B_s\to K\overline{K}$ decays can be measured, 
we will probably have a better understanding of rescattering effects 
anyway, through studies of $B_{u,d}\to\pi K$ and $B^\pm\to K^\pm K$ modes 
at the $B$-factories, which will start operating in the near future. A 
similar comment applies to contributions from electroweak penguins. 

In order to go beyond these constraints and to determine $\gamma$ from the 
$B_{u,d}\to\pi K$ and untagged $B_s\to K\overline{K}$ decays separately, 
the magnitudes of the amplitudes $T$ and $T_s$ have to be fixed, introducing 
hadronic uncertainties into the values of $\gamma$ determined this way. 
Such an input can be avoided by considering the contours arising in the
$\gamma\,$--$\,r_{(s)}$ and $\gamma\,$--$\,\cos\delta_{(s)}$ planes, and 
applying the $SU(3)$ flavour symmetry to relate $r_s$ to $r$ and 
$\cos\delta_s$ to $\cos\delta$, respectively. Using the formalism presented 
in this paper, rescattering and electroweak penguin effects can be included 
in these contours. As a ``by-product'', this strategy yields also values for 
the hadronic quantities $r_{(s)}$ and $\cos\delta_{(s)}$, which are of special
interest to test the factorization hypothesis.

Provided a tagged, time-dependent measurement of the decays $B_s\to K^+K^-$ 
and $B_s\to K^0\overline{K^0}$ can be performed, it would be possible to 
extract $\gamma$ in a way taking into account rescattering effects 
``automatically''. To this end, the $B_s\to K\overline{K}$ observables are
sufficient, i.e.\ no additional input, such as the $SU(3)$ flavour symmetry,
is required, and the theoretical accuracy of $\gamma$ would only be limited 
by electroweak penguins. 

If future experiments should find that the ratio $R_H/R_L$ of the observables
of the untagged $B_s\to K^0\overline{K^0}$ rate is significantly larger 
than ${\cal O}(10\%)$, we would have an indication of physics beyond the 
Standard Model leading to additional CP-violating contributions to 
$B^0_s$--$\overline{B^0_s}$ mixing. The transition $B_s\to D_s^+D_s^-$ 
provides an even more powerful probe of such a scenario of new physics, and
allows a clean determination of the corresponding CP-violating new-physics
phase. Moreover, a comparison of the observables of the untagged 
$B_s\to K^0\overline{K^0}$ and $B_s\to D_s^+D_s^-$ rates may indicate 
sizeable new-physics contributions to the decay amplitude of the former 
channel. If new physics should manifest itself exclusively through a 
CP-violating contribution to $B^0_s$--$\overline{B^0_s}$ mixing, the CKM 
angle $\gamma$ can still be determined with the help of the strategies 
proposed in this paper. Interestingly, the experimental analysis of the 
modes $B_s\to K^+K^-$ and $B_s\to K^0\overline{K^0}$ may even become easier, 
although $|\Delta\Gamma_s|$ is reduced through new-physics contributions 
to $B^0_s$--$\overline{B^0_s}$ mixing. This is because -- in contrast to the 
Standard Model case -- the parts of the untagged $B_s\to K\overline{K}$ 
rates evolving with $e^{-\Gamma_H^{(s)} t}$ are no longer essentially due 
to terms proportional to ${\cal O}(r_s^2)$, ${\cal O}(r_s\,\rho_s)$ and 
${\cal O}(\rho_s^2)$. Keeping only terms linear in $r_s$, $\rho_s$ and 
$\epsilon_s$, $\gamma$ can be extracted by using only the observables of 
the untagged $B_s\to K\overline{K}$ decays as a function of the electroweak 
penguin parameter $\epsilon_s\cos\Delta_s$. This strategy would be 
considerably easier than the one making use of tagged $B_s\to K\overline{K}$ 
decays noted in the previous paragraph.

At present, data for the $B_{u,d}\to\pi K$ modes are already starting to 
become available. On the other hand, time-dependent experimental studies of 
$B_s$ decays are regarded as being very difficult, since in general rapid 
oscillatory $\Delta M_st$-terms have to be resolved. These terms cancel, 
however, in untagged data samples of $B_s$ decays, which have played a major 
role in this paper. Here the width difference $\Delta\Gamma_s$ provides an
interesting tool to extract CKM phases. Such untagged studies are clearly
more promising, in terms of efficiency, acceptance and purity, than tagged 
measurements. However, a lot of statistics is required and hadron machines 
appear to be most promising for such experiments. The feasibility 
depends of course also crucially on a sizeable width difference 
$\Delta\Gamma_s$. Certainly time will tell whether it is large enough to 
constrain -- and eventually extract -- $\gamma$ with the help of the 
$B_s\to K^+K^-$ and $B_s\to K^0\overline{K^0}$ decays discussed in this 
paper. If we are lucky, these modes may even shed light on the physics 
beyond the Standard Model.

\end{document}